\documentclass[natbib,graybox]{svmult}

\usepackage{type1cm}        
\usepackage{makeidx}         
\usepackage{graphicx,amsmath,amssymb}        
\usepackage{multicol}        
\usepackage[bottom]{footmisc}

\usepackage{newtxtext}       %
\usepackage{newtxmath}       
\usepackage{gensymb}         
\usepackage{xcolor}          

\newcommand{\ion}[2]{#1\,{\sc{#2}}}

\makeindex             

\begin{document}

\title*{Hot atmospheres of galaxies, groups, and clusters of galaxies}
\author{Norbert Werner and Fran\c{c}ois Mernier}
\institute{Norbert Werner \at MTA-E\"otv\"os University Lend\"ulet Hot Universe Research Group, P\'azm\'any P\'eter s\'et\'any 1/A, Budapest, 1117, Hungary \\
            Department of Theoretical Physics and Astrophysics, Faculty of Science, Masaryk University, Kotl\'a\v{r}sk\'a 2, Brno, 611 37, Czech Republic \\
            School of Science, Hiroshima University, 1-3-1 Kagamiyama, Higashi-Hiroshima 739-8526, Japan \\
\email{wernernorbi@gmail.com}
\and Fran\c{c}ois Mernier \at MTA-E\"otv\"os University Lend\"ulet Hot Universe Research Group, P\'azm\'any P\'eter s\'et\'any 1/A, Budapest, 1117, Hungary \\
E\"otv\"os University, Institute of Physics, P\'azm\'any P\'eter s\'et\'any 1/A, Budapest, 1117, Hungary \\ 
SRON Netherlands Institute for Space Research, Sorbonnelaan 2, 3584 CA Utrecht, The Netherlands\\
\email{francois.mernier@esa.int}}

\maketitle

\abstract{Most of the ordinary matter in the local Universe has not been converted into stars but resides in a largely unexplored diffuse, hot, X-ray emitting plasma. It pervades the gravitational potentials of massive galaxies, groups and clusters of galaxies, as well as the filaments of the cosmic web. The physics of this hot medium, such as its dynamics, thermodynamics and chemical composition can be studied using X-ray spectroscopy in great detail. Here, we present an overview of the basic properties and discuss the self similarity of the hot ``atmospheres'' permeating the gravitational halos from the scale of galaxies, through groups, to massive clusters. Hot atmospheres are stabilised by the activity of supermassive black holes and, in many ways, they are of key importance for the evolution of their host galaxies. The hot plasma has been significantly enriched in heavy elements by supernovae during the period of maximum star formation activity, probably more than 10 billion years ago. High resolution X-ray spectroscopy just started to be able to probe the dynamics of atmospheric gas and future space observatories will determine the properties of the currently unseen hot diffuse medium throughout the cosmic web. 
}

\begin{figure}
\begin{center}
\includegraphics[scale=0.175]{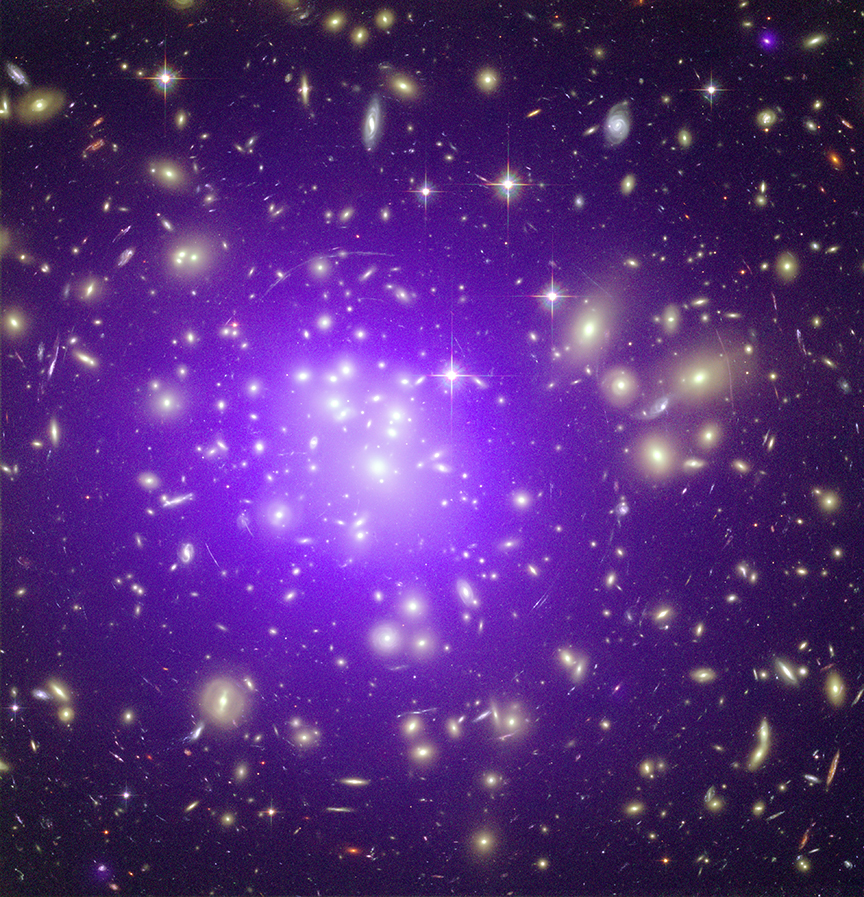} 
\includegraphics[scale=0.182]{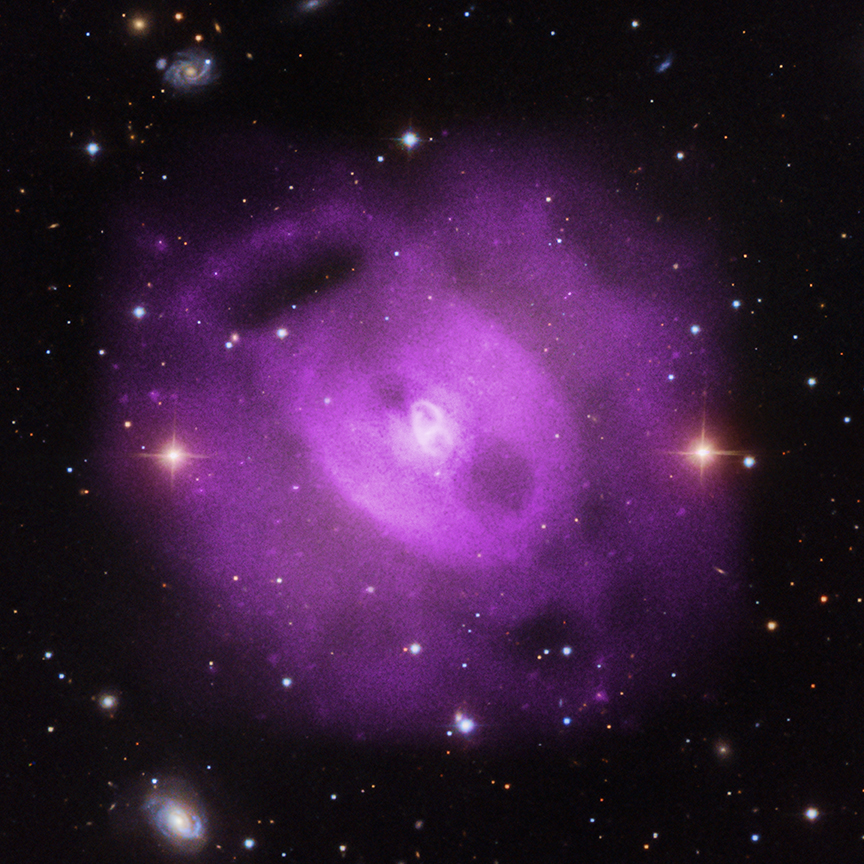}
\caption{These composite optical and X-ray images show the hot, X-ray emitting atmospheres pervading the massive galaxy cluster Abell 1689 (left) and the massive elliptical galaxy NGC~5813 in the centre of a group (right). Courtesy: NASA/CXC/SAO} 
\label{fig:fig1}
\end{center}
\end{figure}

\section{Introduction}
\label{sec:1}

Most of the matter in the Universe remains unseen. The invisible dark matter consists of particles that we have not yet been able to detect. Most of the ordinary matter, which we call baryons, has not been converted into stars and resides in a hot diffuse medium permeating the extended halos of galaxies, 
 groups, and clusters of galaxies, as well as the filaments of the cosmic web. 
 
 In the innermost parts of galaxies the hot, strongly ionised, X-ray emitting plasma is often referred to as {\it interstellar medium (ISM)} and at large radii, well beyond the stellar component, as {\it circumgalactic medium (CGM)}, in groups of galaxies as {\it intragroup medium (IGrM)}, in clusters of galaxies as {\it intracluster medium (ICM)}, and in the filaments of the cosmic web as {\it intergalactic medium (IGM)} or {\it warm-hot intergalactic medium (WHIM)}. Since the hot phases of all these media share crucial similarities - e.g. they can be well described by equations of ideal gas in a gravitational potential - we will refer to them at all scales as {\it hot atmospheres}.

The hot atmospheres permeating massive objects can be probed using X-ray space observatories (see the composite optical/X-ray images of a galaxy cluster and a giant elliptical galaxy  in Fig.~\ref{fig:fig1}). They have been studied in detail in clusters of galaxies, which are particularly bright and can be observed out to redshifts $z=1.5$--2, which corresponds to a look back time of around 10 billion years. However, the crucial role of atmospheres for the formation and evolution of individual massive galaxies is just beginning to be appreciated. They have been discovered using the {\it Einstein} satellite \citep{forman1985} and today, they are studied mainly using {\it Chandra} and {\it XMM-Newton}, which are equipped with CCD-type detectors sensitive in the approximately 0.3--10 keV band, have an energy resolution in the range of 50--150 eV, and a spatial resolution of about 1 arcsec and 15 arcsec, respectively. 

However, outside the dense central regions of massive dark matter halos the atmospheric density becomes too low to be probed by current instruments. Yet, the properties of this mostly unseen diffuse medium are set by large scale shocks and cosmic feedback, therefore its observations with X-ray space telescopes could, in principle, provide key information about the physics of structure and galaxy formation. We therefore expect that the study of hot atmospheres will remain on the frontiers of astrophysics for the decades to come.

\subsection{The simple state of hot atmospheres}

Hot X-ray emitting atmospheres have densities in the range from $\sim10^{-1}$~cm$^{-3}$ in the centres of massive galaxies to $\sim10^{-4}$~cm$^{-3}$ in the outskirts. Their temperatures span from a few $10^6$~K in galaxies to $10^8$~K in the most massive clusters of galaxies. These high temperatures lead to strong ionization. The tenuous hot medium thus consists of fully ionized hydrogen and helium, enriched with highly ionized heavier elements to about a third of their Solar abundances, increasing to around Solar at the centres. This strongly ionized medium has frozen in magnetic fields with strengths of a few $\mu$G. The giroradii of thermal electrons and protons of $\sim10^8$ cm and $\sim10^{10}$ cm (of the order of the size of the Earth), respectively, are many orders of magnitude smaller than the particle collisional mean free paths of $10^{21}-10^{23}$~cm (about 1--100 kpc, comparable to the size of the Galaxy). The ratio of thermal to magnetic pressure $\beta=p/p_{\rm B}\gg1$, which means  that the magnetic pressure contribution is negligible for the hydrostatic mass determination discussed in Sect. \ref{sec:1.1.4}. More generally, even though the hot atmospheres consist of weakly magnetised plasma, equations of ideal gas often provide useful approximations for their properties and in this review, we will for convenience also use the term ``hot gas''. 

At the densities and temperatures of hot atmospheres the ionisation equilibration timescale is short and therefore they are in a state of \textbf{collisional ionisation equilibrium} (CIE). This means that the ionisation rate is equal to the recombination rate, it is determined only by collisions and outside radiation fields do not affect the ionisation state of the plasma. The time scale of the electron-ion equilibration via Coulomb collisions is mostly shorter than $\sim1$ Gyr, therefore we can generally assume that the free electrons are in thermal equilibrium with ions. 
Hot atmospheres are mostly \textbf{optically thin}, which
  means that the density of the gas is so low that the emitted photons will never interact with other ions or electrons within the atmosphere and will thus leave the system in ``straight lines'' \citep[the exception is resonant scattering at the energies of some of the strong emission lines in the dense central regions, as discussed in Sect.~\ref{sec:6}; see][]{churazov2010b}. This, combined with the fact that
the X-ray emissivity is proportional to the square of the gas density (Sect. \ref{sec:1.1.2}), implies that X-ray images of hot atmospheres give us a direct view on the projected gas distribution, considerably facilitating the modelling of spectra and the interpretation of the observations.

\subsection{X-ray emission processes}
\label{sec:1.1.2}

As discussed above, the atmospheric gas is almost entirely ionized, made of ions and free electrons. Under these conditions, a free electron sometimes flies close to an ion and the electrostatic interaction between the two deflects the trajectory of the much less massive electron. The loss of energy due to the deceleration of the electron is then converted into a \textbf{bremsstrahlung} photon that can be detected at X-ray wavelengths. The emissivity $\epsilon$ of this process scales as
\begin{equation}\label{eq:bremsstrahlung}
    \epsilon \propto \bar{g} n_e^2 (kT_\text{e})^{-1/2} e^{-E/kT_\text{e}},
\end{equation}
where $\bar{g}$ is the Gaunt-factor, which is a quantum mechanical correction factor that depends weakly on energy, $k$ is the Boltzmann constant, $T_\text{e}$ is the electron temperature\footnote{By convention, the temperature is often expressed in units of \textit{energy}, as $kT_{\rm e}$, in kiloelectronvolts (keV).}, and $n_{\rm e}$ is the electron density\footnote{The ion density, $n_{\rm i}$, of the plasma with Solar metallicity can be obtained by assuming $n_{\rm e}/n_{\rm i}\simeq 1.18$.} of the gas. Since such interactions can happen at a range of angles and velocities, the energy of the emitted photons can span a range of values. Bremsstrahlung thus contributes to the \textit{continuum} of the X-ray emission. This process is dominant in rich, massive clusters permeated by gas with temperatures $\gtrsim 10^{7.5}$ K (i.e. $kT \gtrsim 3$ keV). 

In addition to hydrogen and helium, hot atmospheres contain traces of highly ionized heavier elements. Like protons, these heavy ions also interact with free electrons, either by capturing them (\textit{radiative recombination}, which also contributes to the continuum emission) or by colliding with them, which often boosts the energy of the bound electrons. These excited electrons eventually return to their lower, initial energy level (\textit{de-excitation}) and, to conserve energy, emit an X-ray photon of an energy corresponding to the difference between the two energy levels in the ion. This produces \textbf{metal emission lines}, each of which corresponds to a specific ion of a specific chemical element. 

\begin{figure}
\begin{center}
\includegraphics[scale=0.65]{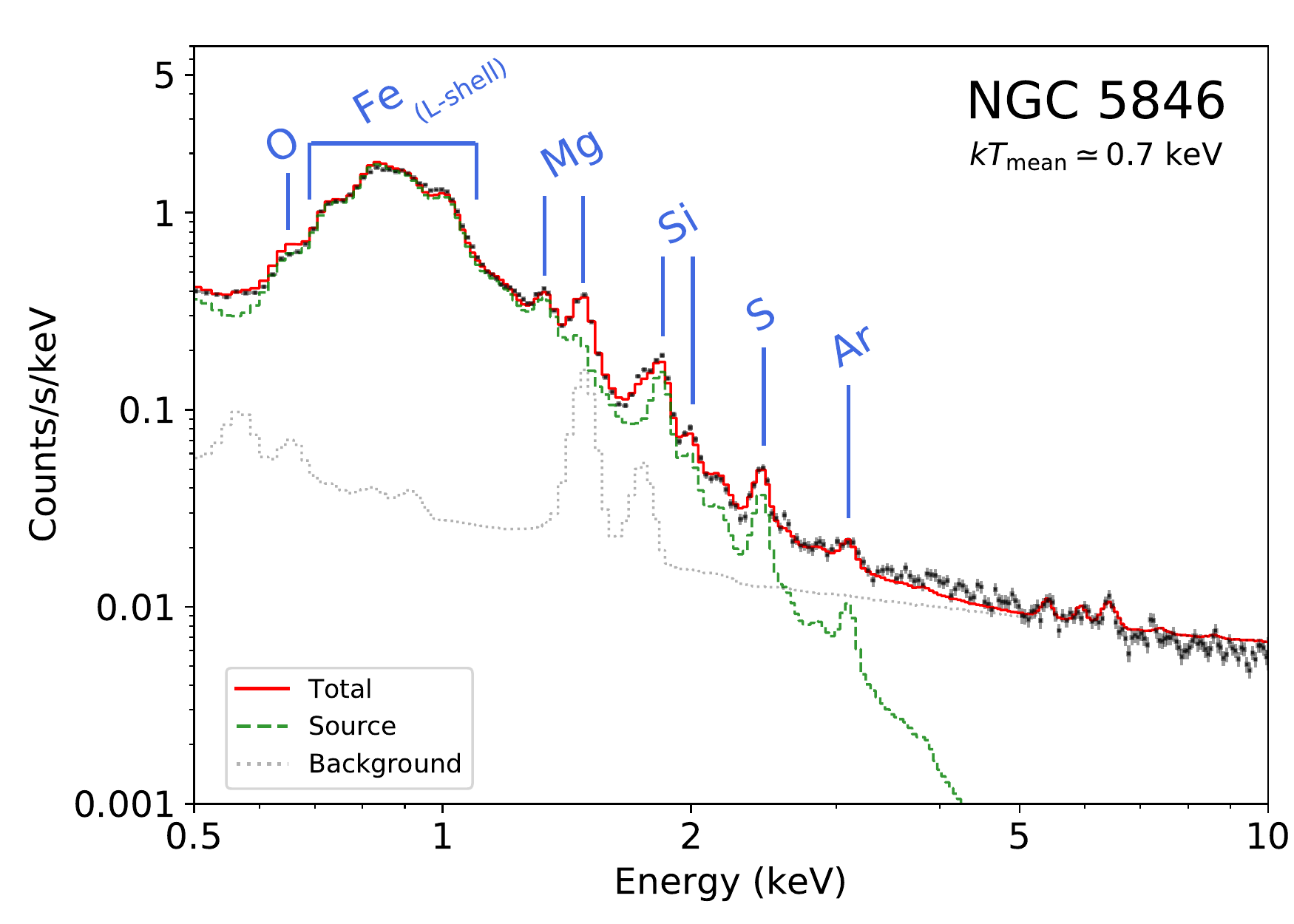} 
\includegraphics[scale=0.65]{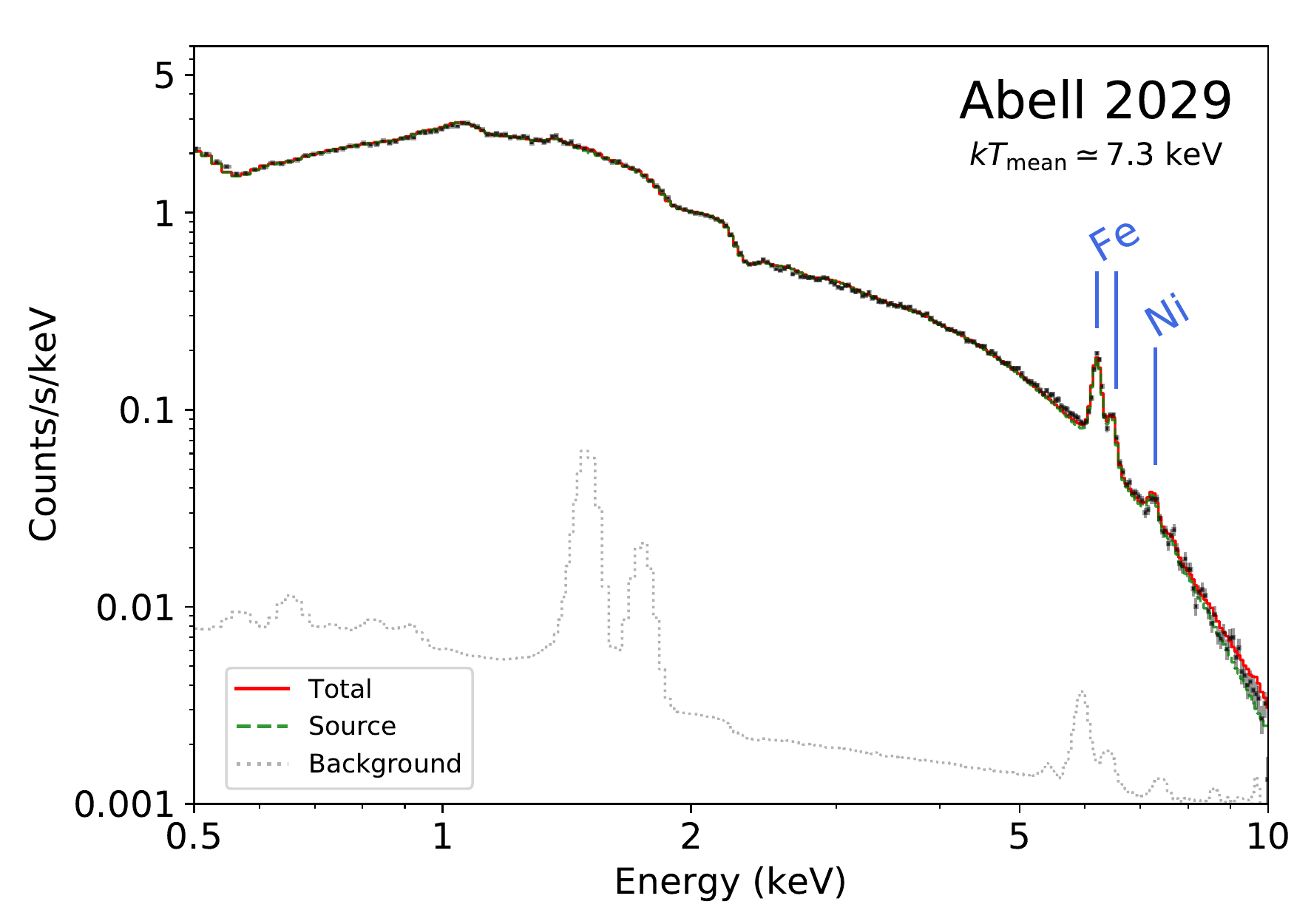} 
\caption{\textit{XMM-Newton}/EPIC MOS spectra of the giant elliptical galaxy NGC\,5846 (top) and the galaxy cluster Abell\,2029 (bottom), both extracted within 0.1$r_{500}$. When visible, metal emission lines (from their K-shell transitions, unless otherwise stated) are labeled. The MOS\,1 and MOS\,2 spectra have been stacked for clarity.}
\label{fig:spectra}
\end{center}
\end{figure}

In cooler gas pervading groups and massive ellipticals ($\sim 10^{6.5}-10^{7.5}$ K, i.e. $kT \sim 0.3-3$ keV), the plasma is less ionized and more of these atomic transitions can occur, giving rise to a large number of emission lines. In fact, in lower-mass systems, line emission often dominates over bremsstrahlung. Fig. \ref{fig:spectra} shows a typical X-ray spectrum produced by a giant elliptical galaxy (NGC\,5846, top panel) and by a massive, hot galaxy cluster (Abell\,2029, bottom panel). While Abell\,2029 is largely dominated by the bremsstrahlung continuum emission with only two visible Fe and Ni line complexes, most of the emission from NGC\,5846 in the ``bump'' between $\sim$0.6 and $\sim$1.2 keV comes from a blend of Fe-L lines that are unresolved by CCD-type detectors.

\subsection{Quantities derived from X-ray spectroscopy}
\label{sec:1.1.3}

X-ray spectroscopy is a powerful tool to study the physics of hot atmospheres. Typically, the observed X-ray spectra are fitted with spectral models of plasma emission with several free parameters. Below, we list the most important quantities determined from X-ray spectra.

\begin{itemize}
    \item The {\bf electron density} of the hot gas is directly proportional to the square root of the X-ray luminosity. From spectral fitting, one can determine the emission measure $Y= \int n_{\rm e} n_{\rm H} \rm{d} V$, where $n_{\rm H}$ is the number density of protons and $V$ is the emitting volume of the source. By using reasonable assumptions about the emitting volume and assuming a smooth gas distribution within this volume, one can easily work out the gas density. 
    \item Eq. \ref{eq:bremsstrahlung} shows that the bremsstrahlung spectrum will exponentially decrease at energies larger than $kT_{\rm e}$. This can be seen clearly in Fig. \ref{fig:spectra}, where the emission of the cooler NGC\,5846 group falls below the background beyond $\sim$3 keV, while the much hotter Abell\,2029 cluster emits photons even at 10 keV. From the observed shape of the continuum, we can thus work out the gas {\bf temperature}. The relative intensities of emission lines provide an additional constraint on the gas temperature, which is particularly useful in the cooler, line dominated systems.  
    \item As mentioned above, each emission line corresponds to an ion of a particular element. Intuitively, the more abundant a given element is, the stronger its associated lines will appear. The \textbf{abundance} of a given element is measured from the equivalent width of its emission line, i.e. the ratio between the flux of the line and that of the continuum at the position of the line. Note, however, that, at a given abundance, the equivalent width of the line is sensitive to the gas temperature. In the relatively cool NGC\,5846 shown in Fig. \ref{fig:spectra}, the unresolved \ion{Fe}{xvii} lines (i.e. with 10 bound electrons) are prominent at around $\sim$0.7-0.9 keV. These lines do not appear in the massive hot Abell\,2029, because its Fe is much more ionized and its spectrum is dominated by \ion{Fe}{xxv} and \ion{Fe}{xxvi} lines (with 2 and 1 bound electron, respectively). 
\end{itemize}{}

From the temperature and density, one can determine the pressure and the entropy of the gas. Since the hot atmospheres are very tenuous and behave almost like an ideal gas, the \textbf{gas pressure} $P$ is simply defined as 
\begin{equation}
    P = \frac{\rho kT}{\mu m_{\rm p}} = n kT \propto n_{\rm e} kT,
\end{equation}{}
where $\rho$ is the gas density, $m_{\rm p}$ is the proton mass and $\mu \simeq 0.6$ is the mean molecular weight given in units of the proton mass.

Another quantity of interest is the {\bf gas entropy}, $K$. In hot atmospheres, it is defined as\footnote{This ``entropy'' is not exactly the same as the classical thermodynamical entropy $s$. Formally, the two quantities are related as $s = k~{\rm ln} K^{3/2} + \text{constant}$.}

\begin{equation}\label{eq:entropy}
    K = \frac{k T}{n_{\rm e}^{2/3}}. 
\end{equation}
In the case of a relaxed cluster in hydrostatic equilibrium, the gas with low entropy naturally sinks to the center while the gas with higher entropy floats toward higher altitudes. If a parcel of gas is moved to a different altitude (e.g. by the activity of a supermassive black hole or by a merger) its temperature and density will change due to adiabatic expansion or compression (assuming that thermal conduction is negligible in the hot plasma), but its entropy will remain unchanged. In this sense, it is a very useful quantity because it keeps memory of the thermodynamic history of the gas. The gas entropy can only be changed by shocks, radiative cooling, conduction, or mixing.

\subsection{Hydrostatic equilibrium}
\label{sec:1.1.4}

The time that it takes for a sound wave to cross an atmosphere provides the timescale on which deviations from hydrostatic equilibrium are evened out. Therefore, if the atmosphere is undisturbed by mergers or violent active galactic nucleus (AGN) outbursts, after several sound crossing times (of order $10^9$ yr for galaxy clusters) hot atmospheres should come to \textbf{hydrostatic equilibrium} with the underlying gravitational potential $\Phi$, meaning that the gravity will be balanced by gas pressure: 
\begin{equation}\label{eq:HE1}
    \frac{1}{\rho} \frac{{\rm d} P}{{\rm d} r} = - \frac{{\rm d} \Phi}{{\rm d} r} = - \frac{G M_\text{tot}}{r^2},
\end{equation}
where $G$ is the gravitational constant and $M_\text{tot}$ is the total (dark matter + stars + gas) mass of the system. This assumption allows us to derive the total mass distribution of the object simply by measuring the basic properties of its X-ray emitting gas. More specifically, one can write:
\begin{equation}\label{eq:HE2}
    M_\text{tot}(<r) = - \frac{kT(r)r}{G \mu m_{\rm p}} \Bigg( \frac{{\rm d}  \ln T}{{\rm d}  \ln r} + \frac{{\rm d} \ln \rho}{{\rm d} \ln r} \Bigg).
\end{equation}{}
As long as the data are good enough to derive the spatial distribution of the gas density and temperature, the total mass encircled within a certain radius, $M_\text{tot}(<r)$, can be readily obtained. 

The departure from hydrostatic equilibrium may be significant in clusters undergoing major mergers, where the gas is heavily disturbed. However, even in systems with prominent gas sloshing and ongoing AGN feedback, like the well known Perseus cluster, the contribution of gas motions to the total pressure support appears to be small (see Sect.\ref{sec:6}).

\section{Self similarity}
\label{sec:2}

In the $\Lambda$CDM (cosmological constant and cold dark matter) model -- favoured by cosmologists today -- clusters and groups of galaxies formed at the density peaks that were born out of quantum density fluctuations in the very early Universe, which later collapsed under their own gravity. These systems grew further in a hierarchical way, i.e. larger structures formed from smaller ones, mainly via steady accretion and mergers.

This paradigm is particularly interesting in the context of the so-called ``self-similarity''. As shown by \citet{kaiser1986}, the assumption that the Universe has the critical density ($\Omega = 1$) implies that the wave-number (i.e. the inverse of the typical spatial scales) of the density fluctuations is distributed as a power-law. The  consequence of this prediction is that, on sufficiently large scales, the Universe is expected to be self-similar. In other words, smaller dark-matter structures - containing massive galaxies or galaxy groups - appear as scaled-down versions of the largest haloes associated with massive galaxy clusters. 

As shown by numerical simulations \citep[e.g.][]{navarro1995}, self-similarity is expected to hold not only for the dark-matter component, but also for the hot, X-ray emitting atmospheres pervading these large scale structures. The gas falling into dark matter halos is expected to be heated by shocks. The more massive a halo is, the more shock-heated, and thus the hotter, the gas becomes. Combining the self-similar assumption with the approximation of hydrostatic equilibrium discussed earlier, one can easily show \citep[see][]{giodini2013} that, within a given overdensity limit\footnote{$r_\Delta$ (with $\Delta = 200$ or $\Delta = 500$ as commonly found in the literature) corresponds to a radius within which the total matter density reaches $\Delta$ times the critical density of the Universe at the redshift of the system. Defined this way, $r_\Delta$ can be associated with a ``normalized'' astrophysical radius common to each system, which takes self-similarity into account.} $r_\Delta$, the total mass $M_\Delta$ of a system scales with its gas temperature $T_\text{$\Delta$}$ as
\begin{equation}\label{eq:1}
    M_\Delta \propto T_\text{$\Delta$}^{3/2} .
\end{equation}{}
This simple relation shows that, in the absence of other mechanisms than gravity and shock heating, the temperature of a system reflects its gravitational potential.

Another important prediction of the self-similar paradigm is that the dimensionless properties of massive galaxies, groups, and clusters are expected to be invariant. In particular, the ``gas mass fraction'', i.e. the fraction of hot, X-ray emitting gas mass over the total (mostly dark matter) mass is expected to be the same in all massive systems. Combining this prediction with the assumption that the total emissivity of the plasma is dominated by thermal bremsstrahlung\footnote{This assumption holds relatively well in the case of hot, massive clusters. However, as we have seen earlier, metal lines usually dominate the emissivity of cooler atmospheres pervading groups and ellipticals, which adds further complications to testing the predictions of self-similarity.} (see Sect. \ref{sec:1}), one can also easily demonstrate that the X-ray luminosity $L_\text{X, $\Delta$}$ of a system scales with its gas temperature as
\begin{equation}\label{eq:2}
    L_\text{X, $\Delta$} \propto T_\text{$\Delta$}^{2} .
\end{equation}{}

\begin{figure}
\begin{center}
\includegraphics[scale=0.46]{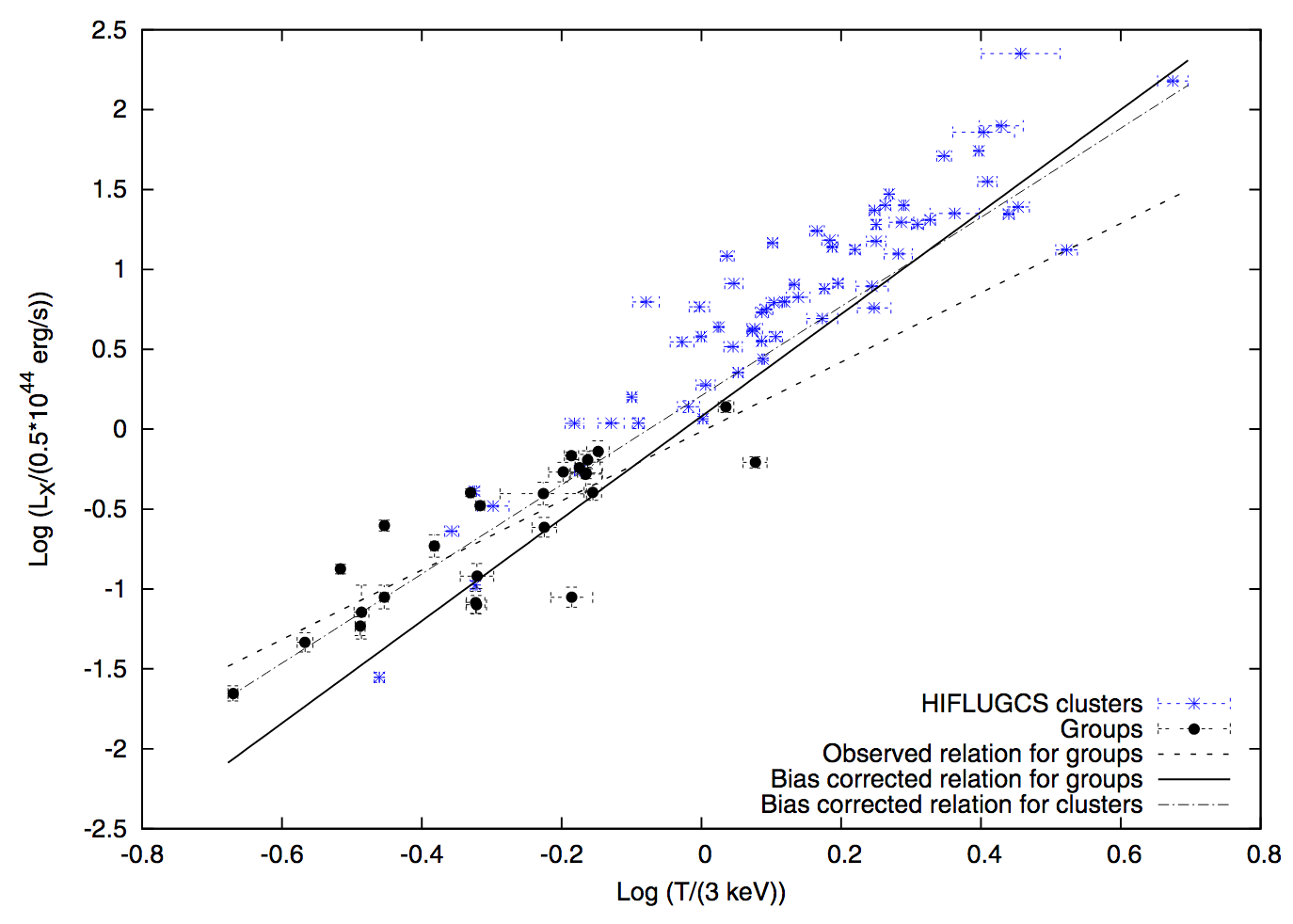}
\caption{X-ray luminosity vs. temperature relation for a sample of clusters and groups \citep{bharadwaj2015}. The solid line and dash-dotted line show the best-fit relations for the sub-samples of groups and clusters, respectively, after corrections for biases due to selection effects. Groups appear to have a steeper relation than clusters.}
\label{fig:L-T}
\end{center}
\end{figure}

Since $L_\text{X, $\Delta$}$ and $T_\text{$\Delta$}$ can be obtained independently from X-ray data (Sect. \ref{sec:1.1.3}), Eq. (\ref{eq:2}) is extremely useful to directly test the self-similar assumption via X-ray observations of clusters, groups, and ellipticals. An example of such a study is shown in Fig. \ref{fig:L-T} \citep[from][]{bharadwaj2015}, however, we stress that many other studies are available in the literature \citep[e.g.][]{zhang2008,pratt2009,eckmiller2011,mahdavi2013,lovisari2015}. 

The general conclusion of these results is that, despite a strong correlation between temperature and luminosity, the slope of the power-law is closer to 3 than to 2, which suggests a deviation from self-similarity. Moreover, as seen in Fig. \ref{fig:L-T}, the relation for groups is steeper than that for clusters.  In other words, groups lay on average below the expected scaling relation, which means that they must be either \textit{hotter} or \textit{less luminous} (or both) than initially expected. Since self-similarity assumes that gravity and shock heating are the only processes that shape the atmospheres of massive systems, the most likely cause for such deviations are additional \textit{non-gravitational} processes, such as energy input by AGN and supernovae.

\begin{figure}[t]
\begin{center}
\includegraphics[scale=0.5]{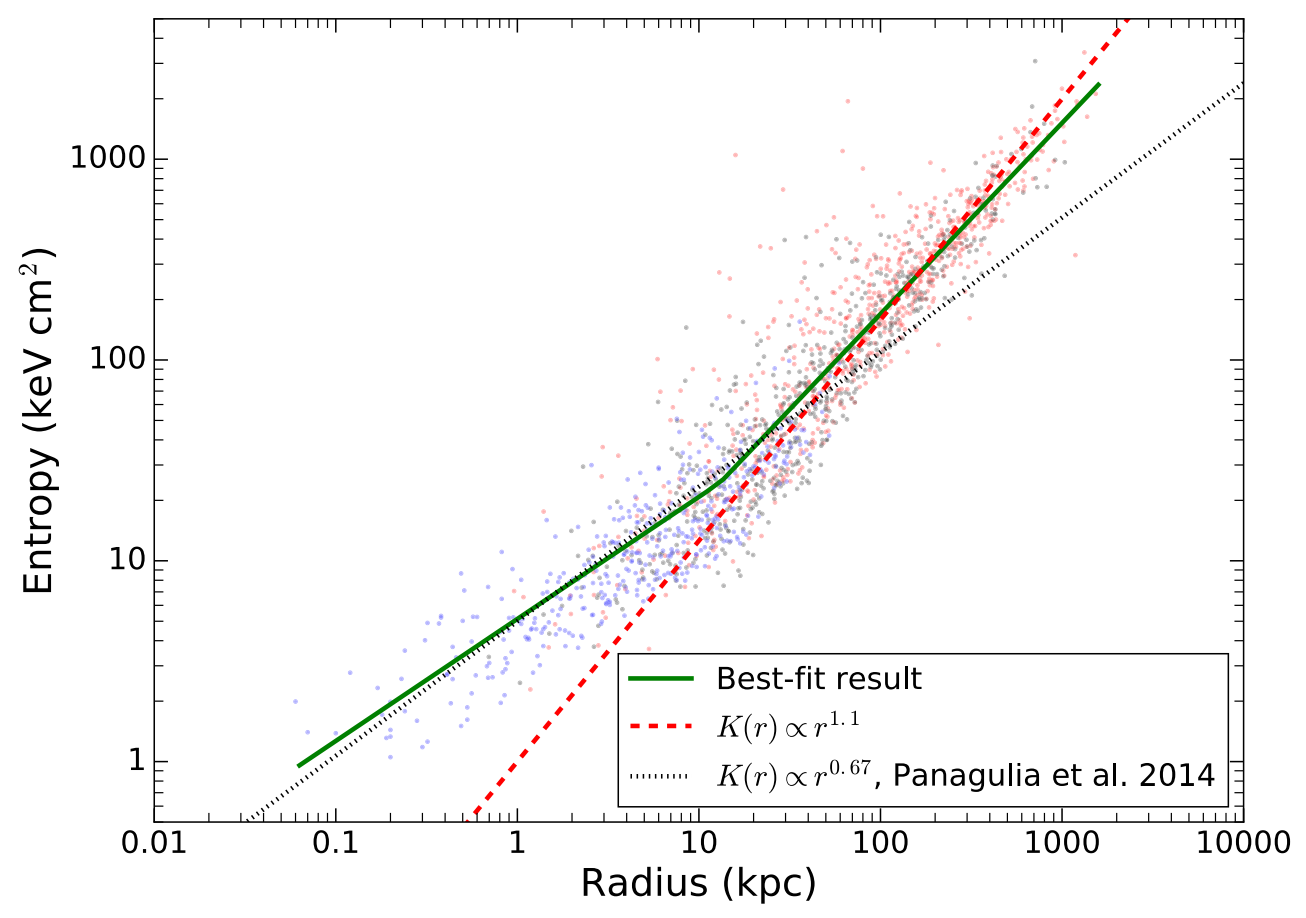}
\caption{Measured radial entropy profiles for elliptical galaxies (blue data points) and for more massive systems (red and black data points). The red dashed line indicates a power-law distribution with index 1.1 predicted by gravitational collapse models and the black doted line indicates a power-law model with index 0.67 found at $r\lesssim50$~kpc by \citet{panagoulia2014}. The data and the best-fit broken power-law model (green solid line) show that the central entropy significantly flattens in the cores of these systems \citep[][and references therein]{babyk2018}.} 
\label{fig:entropy}
\end{center}
\end{figure}

Following its definition, the entropy (Eq. \ref{eq:entropy}) scales linearly with the gas temperature. It has also been shown, using hydrodynamic simulations of gravitationally collapsed gas in hydrostatic equilibrium \citep{tozzi2001,voit2005}, that the radial distribution of entropy is expected to increase with radius following a power-law distribution of 
\begin{equation}
    K(r) \propto r^{1.1} .
\end{equation}{}
This prediction has been tested using X-ray observations of clusters, groups, and ellipticals, as shown in Fig. \ref{fig:entropy}. Compared to the self-similar expectations, it appears that the radial distribution shows an excess of entropy in cluster cores, which is on average stronger in low-mass systems. 

This entropy excess can be explained by a centrally increased temperature and/or by a central gas density drop (see Eq.~\ref{eq:entropy}). In other words, there must be a mechanism at the center of these hot atmospheres that heats the gas and/or moves a significant fraction of the gas up to larger altitudes. The most likely mechanism is the feedback from the central supermassive black hole, which has more impact in the lower mass groups than in massive clusters. It has also been suggested that the entropy excess may be a signature of ``pre-heating'', i.e. a heating of the gas, by AGNs or stellar activity, before the formation of clusters and groups \citep{ponman1999}. 

The idea that a significant fraction of gas has been lifted from the central regions of lower mass systems also agrees with the finding that, at scaled radii of $r_{2500}$ or even at $r_{500}$, the gas mass fraction is an increasing function of cluster mass up to massive clusters with virial temperatures of $kT\sim5$~keV \citep{vikhlinin2006, gastaldello2007,pratt2009,dai2010}. In dynamically relaxed high mass clusters with $kT\gtrsim5$~keV the gas mass fraction at $r_{2500}$ appears to be invariant \citep{allen2008}. 

\section{Hot atmospheres of galaxies and the circumgalactic medium}
\label{sec:4}

\begin{figure}
\begin{center}
\includegraphics[scale=0.23]{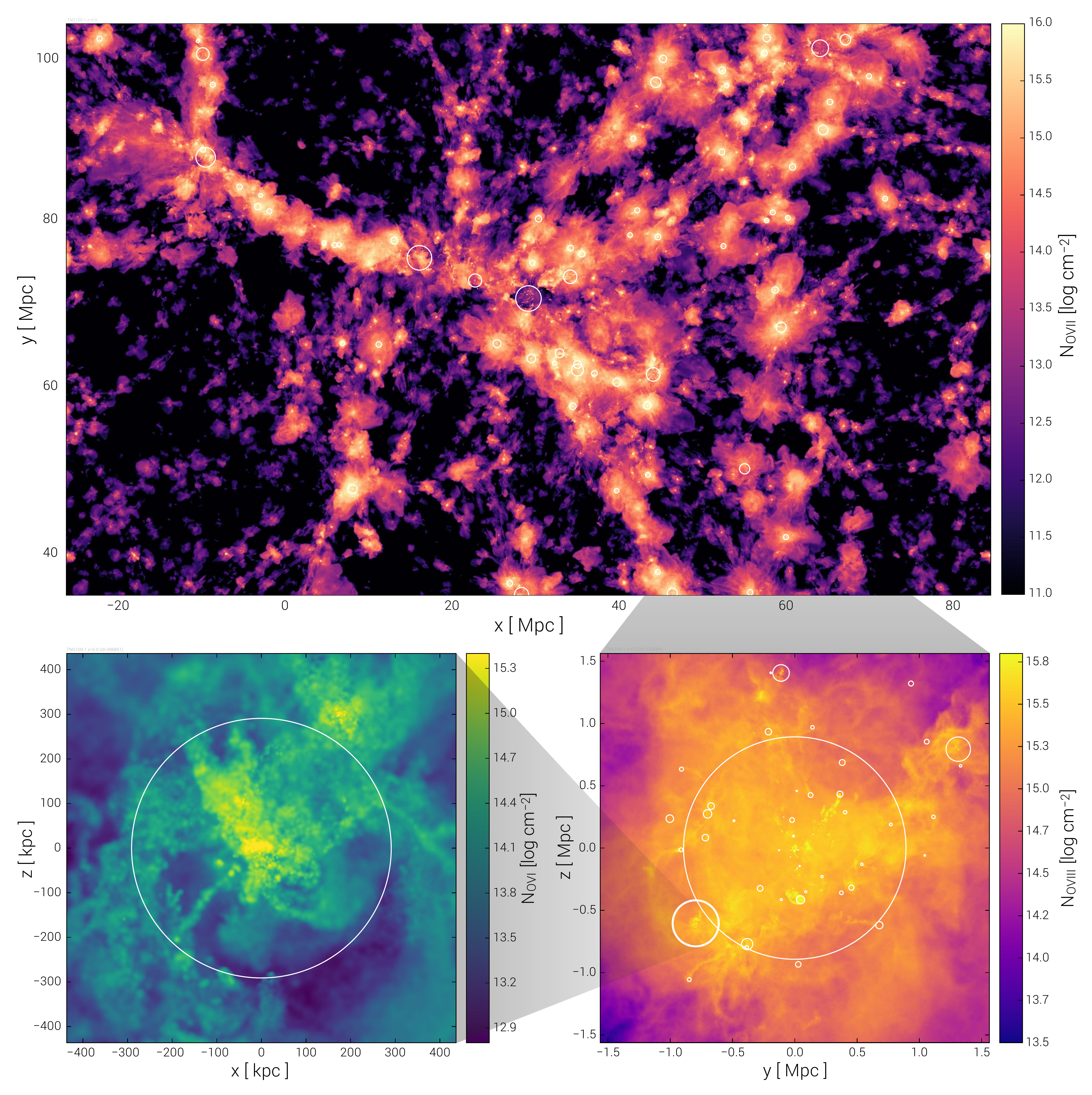}
\caption{These plots show the distribution of highly ionized oxygen, which is an excellent tracer of the warm-hot diffuse gas, in the Illustris TNG100 simulations at $z=0$. The bulk of the oxygen in the circumgalactic medium is expected to reside in \ion{O}{vii} and \ion{O}{viii} ions, which have strong X-ray emission lines. The upper panel shows the column density of \ion{O}{vii} over a depth of 15 Mpc. \ion{O}{vii} traces the large scale filaments of the cosmic web and galaxy sized collapsed halos. The white circles show the 50 most massive halos. The lower right panel zooms in on a galaxy cluster with a halo mass of $10^{13.8}~M_{\odot}$, showing its \ion{O}{viii} column in a box with a length of $3.5~r_{\rm vir}$. The large circle shows the virial radius and the small circles indicate the 50 most massive halos in the panel. The lower left panel zooms in further on a system with a halo mass of $10^{12.5}~M_\odot$, showing the column density of \ion{O}{vi}. From \citet{nelson2018}.}
\label{fig:cosmicweb}
\end{center}
\end{figure}

At least a part of the hot atmospheric gas of galaxies more massive than $M_{\rm crit}\approx10^{12}~M_{\odot}$ \citep[e.g.][]{correa2018} was accreted externally
and shock heated during the process of galaxy assembly. The atmospheres were augmented significantly by
stellar mass loss \citep{pellegrini2018}. The contribution of stellar mass loss material
increases and might actually dominate in the lower mass
systems. Such atmospheric gas contains a large fraction of baryons in the local
Universe (see Fig.~\ref{fig:cosmicweb}). It turns out that about half of the warm-hot diffuse baryons at low redshifts may lie in galactic atmospheres
\citep[e.g.][]{fukugita1998,keres2005,fukugita2006}.

Historically, the extended, volume-filling, hot atmosphere of our Milky Way was revealed by its soft X-ray emission
\citep{snowden1997,henley2010,henley2012} and further probed by X-ray
absorption studies along sightliness to bright AGN
\citep{paerels2003,gupta2012}. These studies show that it contains  $2.5\pm1\times10^{10}~M_{\odot}$ of gas, which is less than its stellar mass \citep{Bland-Hawthorn2016}. The baryonic mass fraction within the virial radius of our Galaxy is only $6\pm1$\%, which falls well short of the Universal cosmic value of 16\%. Studies of other spiral galaxies
with {\it Chandra} and {\it XMM-Newton} find similar conclusions. When
the observed density profiles are extrapolated to the virial radii of
these galaxies, more than 60--70\% of the baryons appear to be missing
\citep{anderson2016,bogdan2017,li2018}. Stacked Sunyaev-Zeldovich (SZ) measurements of massive galaxies by the
{\it Planck} satellite indicates that most of their baryons reside in hot diffuse atmospheres and extend beyond their virial radii
\citep{planck2013,greco2015}. \citet{bregman2018} show that the
atmospheric density profiles of massive spirals need to be
extrapolated to 1.9--3 $R_{200}$ for their baryon to dark matter ratio
to approach the cosmic value.  Some kind of violent activity likely caused these galaxies to expel a large part of their hot atmospheres.

Our estimates of the baryon fractions also depend strongly on the metallicity of the gas, which was previously often assumed low. Contrary to such earlier results \citep[e.g.][]{rasmussen2009,bregman2010,sun2012,yates2017}, the hot galactic atmospheres of early type galaxies are likely metal rich, with metallicities approaching the Solar value in the central regions \citep{mernier2018a} and flattening out at $\approx0.2-0.3$~Solar at larger radii \citep[see][and Sect.~\ref{sec:5}]{mernier2017}. Detailed spatially resolved measurements of metallicities in the hot atmospheres of late type galaxies have not yet been performed. Such measurements are challenging due to the low number of photons. 

Current measurements of the properties of galactic atmospheres are mostly limited to early type galaxies, where the presence of hot atmospheric gas in combination with radio mode AGN feedback is believed to have played a crucial role in the quenching of star formation (see Sect.~\ref{sec:3}).  
However, the detailed knowledge of the atmospheric gas is in many ways essential also for our understanding of the stellar components of late type, discy galaxies. 

As an example, the hot atmospheric gas might also be of key importance for the
evolution of spiral galaxies, feeding their discs with the fuel
necessary for star formation. Milky Way mass spiral galaxies, that are
forming stars at a rate similar to that of our Galaxy ($\approx2~M_\odot$~yr$^{-1}$) would consume
the available molecular gas in their discs in about $10^9$ years. To
maintain the star forming galactic discs, they have to be fed
continuously by molecular gas from outside \citep{fraternali2012}.
Thermal instabilities in the hot atmospheres, leading to a ``rain'' onto the galactic disc (see Sect. \ref{sec:3}), would be able to provide plenty of fuel to maintain the star formation in spirals for $\approx 10^{10}$ years. The cooling of the galactic atmospheres onto star forming discs could be stimulated by stellar feedback via powering of a galactic fountain, which produces mixing between the disc material and the atmospheric gas. The mixing reduces the cooling time of the atmosphere, making it condense and accrete onto the disc \citep{fraternali2017}. In this fountain driven accretion scenario, the key ingredient is the presence of a star-forming disc of cold gas, which helps to cool the atmosphere. If the disc gets for some reason destroyed (e.g. due to ram pressure stripping after falling into a galaxy cluster) it may not reform again
and the galaxy will become red and dead.

Most X-ray observations only target the bright central parts of galaxies, groups and clusters of galaxies. Most of their hot atmospheres, which contains the dominant fraction of their baryons are still unobserved. 
Few systems have reliable observations all the way to their virial radii and we will have to wait for future observatories to really address the baryon content, dynamics, thermodynamics, and the chemical composition of the full extent of their atmospheres (see Sect.~\ref{sec:6}). The existing observations indicate that the gas in the cluster outskirts is clumpy and it is not entirely virialized \citep{walker2019}. The same might be true for the outskirts of individual galaxies, where most of the ordinary matter resides.

Even fewer observations probe the properties of the warm-hot intergalactic medium (WHIM) outside of galaxies, groups, and clusters of galaxies, which is believed to be permeating the filaments of the cosmic web. A few tentative detections in emission are likely probing the densest and hottest parts of the intergalactic medium, where only a small fraction of the WHIM resides \citep[e.g.][]{werner2008a,bulbul2016,alvarez2018}. On the other hand absorption studies provided line detections only in extremely long observations toward a few bright quasars \citep[e.g.][]{nicastro2018,kovacs2019}. A systematic study probing the bulk of the diffuse medium in the outskirts of galaxies and permeating the cosmic web will need a substantial improvement in our observing capabilities (see Sect.~\ref{sec:7}).

\section{Supermassive black hole feedback}
\label{sec:3}

\begin{figure}[t]
\begin{center}
\vspace{0cm}
\includegraphics[scale=0.16]{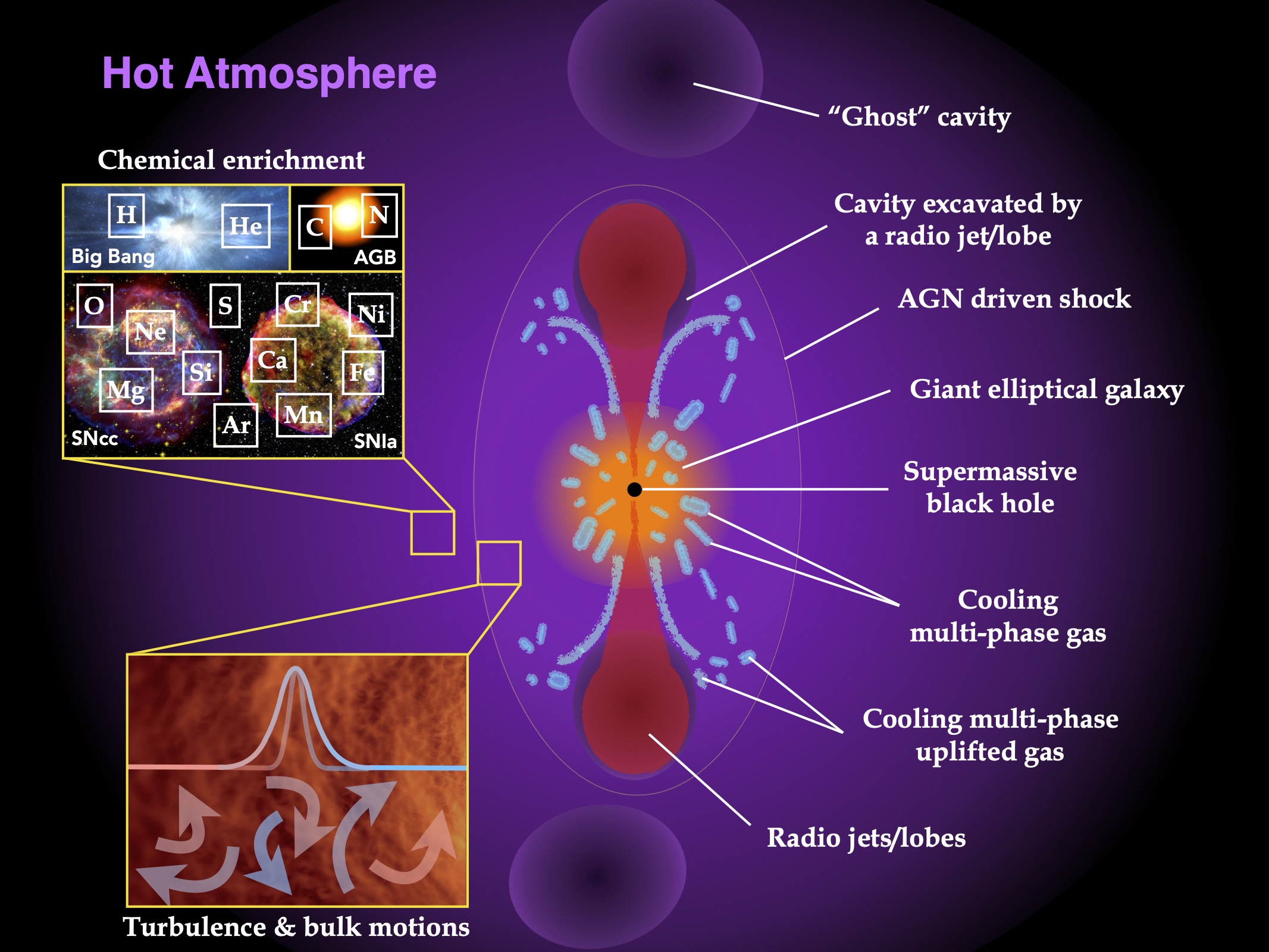}
\vspace{0cm}
\caption{An illustration showing a chemically enriched (by SNIa, SNcc, and AGB stars), hot, X-ray emitting atmosphere, stabilized by AGN feedback. Within a radius where the cooling time of the atmospheric gas is shorter than $\sim1$~Gyr the thermally unstable atmospheric gas ``precipitates'' by condensing into cooler clouds that may fall towards the centre, increasing the accretion rate onto the central supermassive black hole and driving the formation of jets. The jets inflate lobes, displacing the hot atmospheric gas and creating X-ray dark cavities. Initially, the lobes/cavities expand supersonically, driving weak shocks into the surrounding medium and increasing its entropy. After detaching from the jets, the lobes/cavities rise buoyantly in the hot atmosphere, driving turbulence and uplifting low entropy gas in their updraft. The uplifted low entropy gas may cool and fall back towards the centre. As the relativistic plasma filling the bubbles looses energy it stops shining in the radio band and the cavities become ``ghost'' cavities. By mechanically perturbing the gas, e.g. driving shocks, uplifting the lowest entropy gas from the centre, driving turbulence, the AGN heats the atmosphere preventing its radiative cooling. AGN driven turbulent motions cause Doppler broadening of spectral lines, which can be observed by high-resolution X-ray spectrometers.}  
\label{fig:bubbles}
\end{center}

\end{figure}
It is now well established that central supermassive black holes play a vital role in the evolution of their host systems. Accreting black holes are the most efficient engines in the Universe at converting rest-mass into energy, releasing $\sim 10^{20}$~erg per gram of accreted gas. This energy may be released in either \textit{radiative} or \textit{mechanical} form, depending on the accretion rate and structure of the accretion flow: in the radiative (quasar mode) AGN feedback, the radiation from vigorously accreting black holes couples with the cold gas in the host galaxy; in the so-called radio-mechanical (maintenance mode) feedback, the jets from AGN accreting at a modest rate heat or push the hot gas out. Radiatively efficient accretion and rapid black hole growth are taking place mostly early in the evolution of galaxies.

In massive galaxies, groups and clusters of galaxies hosting hot atmospheric gas, the quasar phase is followed by radiatively inefficient radio-mechanical feedback, when despite a much lower accretion rate the mechanical energy output of the jets couples efficiently to the atmospheric gas \citep[e.g.][]{churazov2005}. In the absence of heating, the hot X-ray emitting gas in many clusters, groups and elliptical galaxies would cool and form stars, building much larger galaxies than are seen. However, the radio-mechanical feedback appears to be preventing the cooling of the hot gaseous atmospheres \citep[see][]{mcnamara2012}. 

\begin{figure}[t]
\begin{center}
\vspace{0cm}
\includegraphics[scale=0.33]{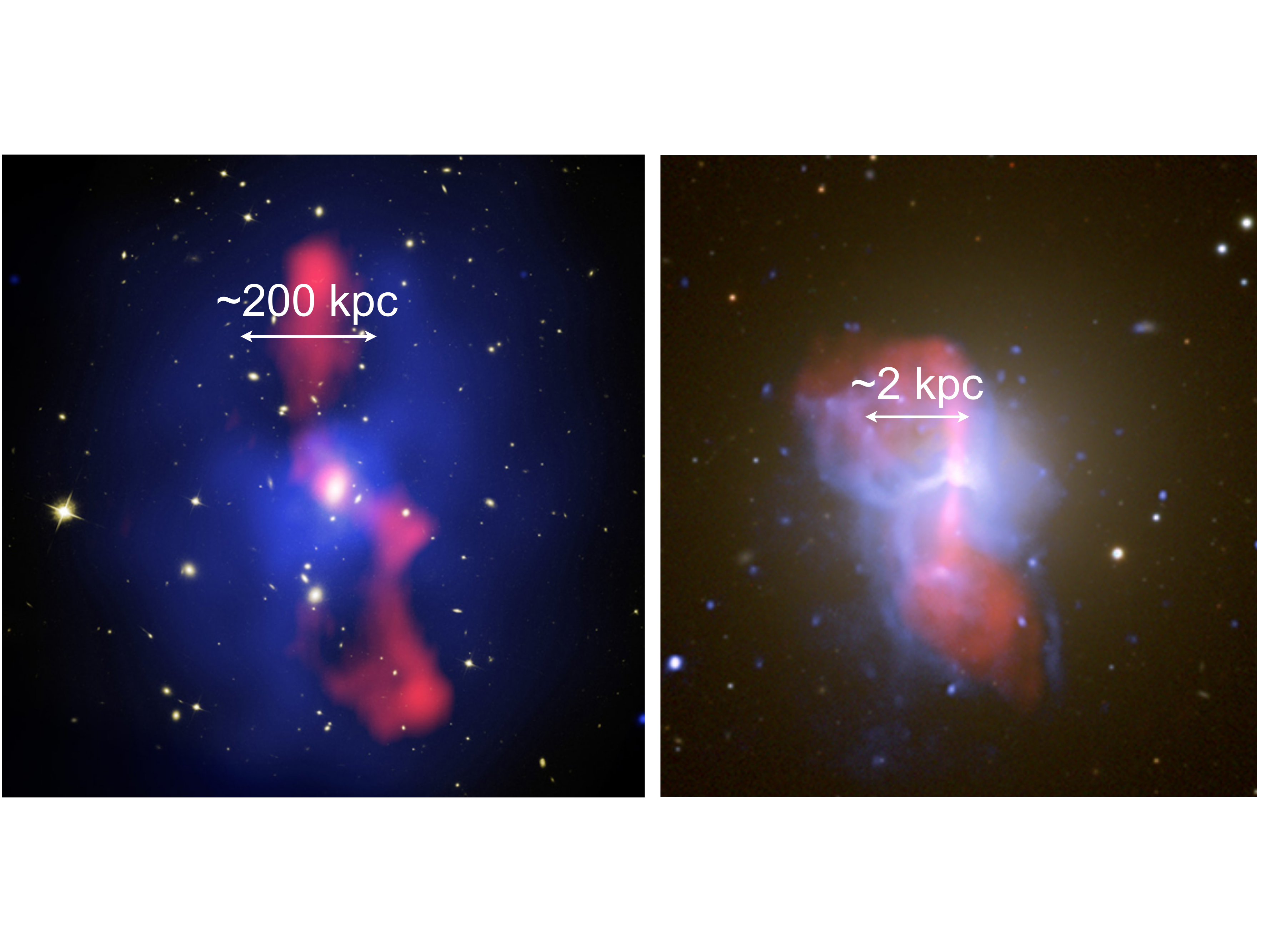}
\vspace{-1cm}
\caption{Radio-mechanical AGN feedback observed on a cluster scale (left panel) and in a galaxy (right panel). {\it Left panel:} Composite image of the galaxy cluster MS 0735.6+7421 \citep{mcnamara2005}. {\it Right panel:} Composite image of M 84, a massive elliptical galaxy in the Virgo cluster \citep{finoguenov2008}. The {\it Chandra} image of the hot X-ray emitting gas is shown in blue and the VLA radio image of the jet injected relativistic plasma is shown in red on both images. A background image from the Sloan Digital Sky Survey is shown in yellow and white. The jet inflated bubbles displace the X-ray emitting plasma, forming  cavities in the hot atmospheres. While the average jet power required to form the observed cavities in MS 0735.6+7421 is $1.7\times10^{46}$~erg~s$^{-1}$ (typical for quasar luminosity but mostly released as mechanical energy), comparable to the X-ray luminosity of this massive cluster, the jet power of M~84 is five orders of magnitude smaller, apparently tuned to balance the radiative cooling of its much smaller X-ray atmosphere. Courtesy: NASA/CXC/SAO} 
\label{fig:bubbles2}
\end{center}
\end{figure}

In the centres of clusters of galaxies, radio-mechanical feedback is relatively well explored. X-ray studies with {\it Chandra} and {\it XMM-Newton} have shown that the expanding, AGN jet inflated radio lobes displace the hot gas, creating `cavities' in the X-ray emitting plasma \citep[e.g.][]{mcnamara2005} and drive weak shocks that heat the surrounding medium isotropically \citep{nulsen2005,forman2005,forman2007,simionescu2009b,million2010b}, preventing it from cooling (see Fig. \ref{fig:bubbles} and \ref{fig:bubbles2}). Furthermore, in their wakes, the rising bubbles filled by relativistic plasma uplift low entropy gas from the innermost regions of their host galaxies \citep{simionescu2008, simionescu2009a,werner2010,werner2011,kirkpatrick2011,kirkpatrick2015}. Current X-ray observations thus show that jets emanating from black holes accreting at modest rates are sufficiently powerful to balance the radiative cooling of hot atmospheres and limit further star-formation. The feedback is gentle and the heating rate appears to be very well tuned to the atmospheric cooling (see Fig.~\ref{fig:bubbles2}). However, no consensus has been achieved on the dominant mechanism responsible for energy transport from jets, to X-ray bubbles, and eventually into the hot plasma at large \citep[see][]{werner2019}.

There is also little consensus on how cooled or cooling gas fuels the AGN. To create a feedback loop, the thermal state of the hot atmosphere must influence the power output of the AGN. A particularly important question, which still remains unanswered is: {\it How does AGN feedback operate across spatial scales of over 8 orders of magnitude - from the immediate vicinity of the black hole to a scale of several hundreds of kpc in clusters of galaxies? }

The most popular attempts to solve this question rely on the idea of ``precipitation'' \citep{gaspari2013,voit2015c,mcnamara2016}. If the ambient medium ``precipitates'' (cools) by condensing into cooler clouds that rain toward the centre, the accretion rate will rise by orders of magnitude and trigger a feedback response.

Supermassive black hole masses correlate with the luminosities and velocity dispersions of the stellar bulge components of their host galaxies indicating that they evolved together \citep{Magorrian1998,Gebhardt2000,FerrareseMerritt2000,KormendyHo2013,Saglia2016}. Recently, \citet{lakhchaura2019} found a tight correlation between the central supermassive black hole mass and the atmospheric temperatures of the brightest cluster/group galaxies (BCGs). The atmospheric temperatures are determined by the total mass of the system, implying an underlying correlation between the black hole mass and the total mass of the host galaxy. The hydrostatic analysis of BCGs confirms the existence of the underlying correlation, which turns out to be approximately linear. These results imply that the black hole masses correlate not only with the stellar components of the galaxies but also with their dark matter halos \citep[see also the earlier results of][]{bogdan2015, bogdan2018}. The supermassive black hole mass could be determined by the binding energy of the halo through radiative feedback during the rapid black hole growth by accretion \citep{booth2010,booth2011}. However, for  the most massive galaxies, such as the BCGs, mergers are the chief channel of growth \citep{weinberger2018}. Thus the initial correlation established through quasar mode feedback could have been subsequently strengthened via numerous gas-free mergers with galaxies hosting central supermassive black holes \citep[a natural consequence of the central limit theorem;][]{jahnke2011}.

\section{Atmospheric gas dynamics}
\label{sec:6}

Next to the standard X-ray observables, such as the density, temperature, and metallicity discussed above, X-ray detectors with high spectral resolution also enable direct measurements of the hot gas dynamics. While the broadening of spectral lines provides information about small scale or turbulent motions, line shifts reveal bulk, coherent, ``streaming'' motions on larger scales. Infall of groups and galaxies as well as accretion of surrounding matter, and the interaction of the intra-cluster medium (ICM) with the relativistic radio emitting plasma injected by the AGN are all expected to drive turbulent and bulk gas motions. Actual measurements of gas motions, however, remain sparse. The knowledge of the dynamical pressure support is also very important for the total mass measurements performed under the assumption of hydrostatic equilibrium and for cluster cosmology, which relies on accurate masses.

\begin{figure}[t]
\includegraphics[scale=.33]{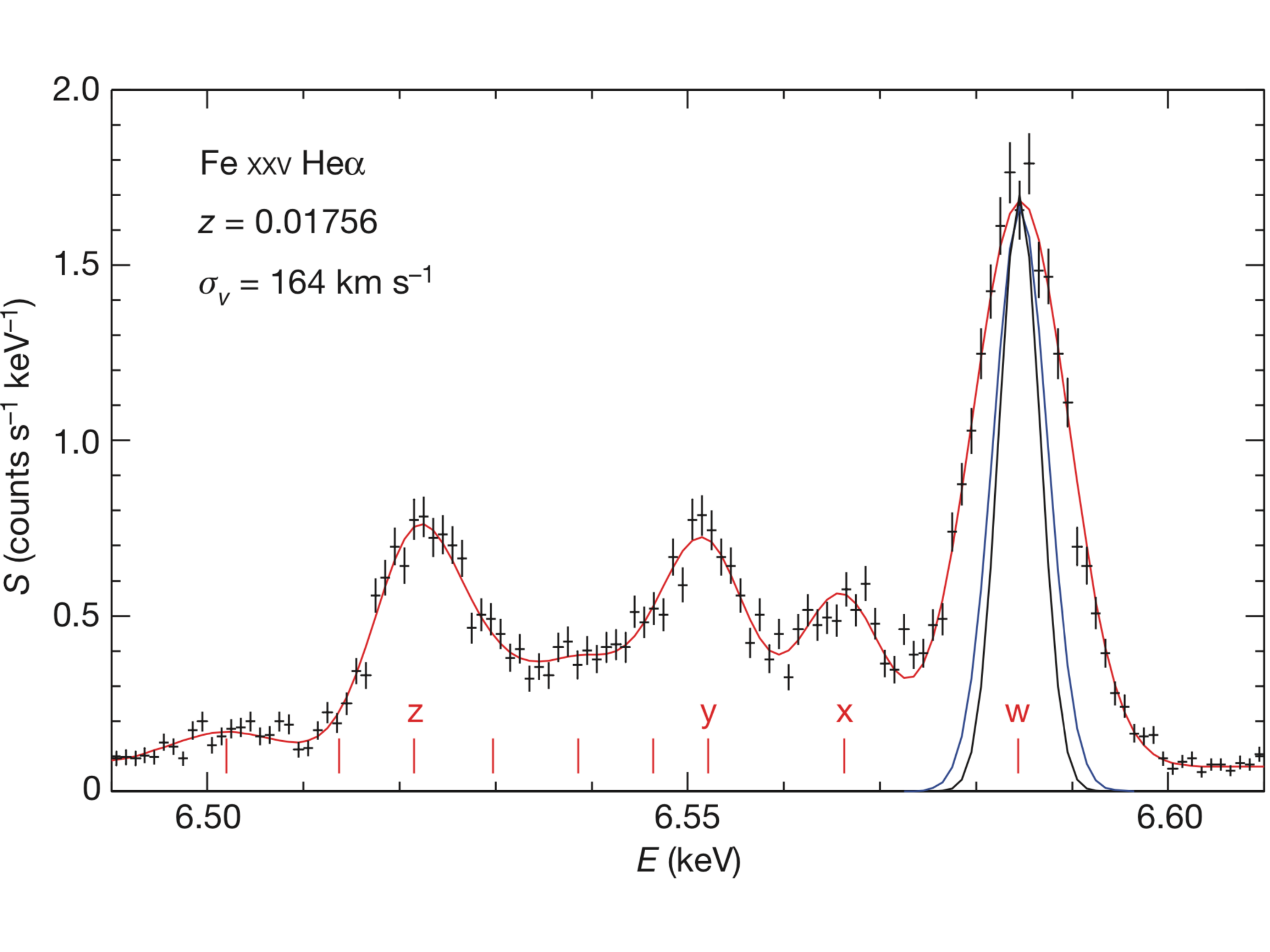}
\caption{The {\it Hitomi} SXS spectrum of \ion{Fe}{xxv} He$\alpha$. The black line indicates the instrumental broadening, the blue line indicates the instrumental plus thermal broadening. All the additional broadening - the difference between the blue line and the data fitted by the red model - is due to velocity broadening. The strongest resonance (`w'), intercombination (`x', `y') and forbidden (`z') lines are indicated \citep{hitomi2016}.}
\label{linebroad}       
\end{figure}

The first non-dispersive imaging detector with a high enough spectral resolution to determine the velocity broadening and shifts of ICM emission lines was the Soft X-ray Spectrometer (SXS) on the {\it Hitomi} satellite (see Fig.~\ref{linebroad}). The  {\it Hitomi}  SXS  observation of the Perseus cluster revealed a 1D line-of-sight velocity dispersion reaching approximately 200~km~s$^{-1}$ toward the central AGN and toward the jet inflated ``ghost'' bubble north of the cluster core \citep{hitomi2016,hitomi2018vel}. Elsewhere in the centre of the Perseus cluster probed by the {\it Hitomi} observations, the velocity dispersion of the ICM appears constant at around 100~km~s$^{-1}$. The observations also revealed a velocity gradient with a 100~km~s$^{-1}$ amplitude across the cluster core, consistent with a large-scale sloshing of the core gas. It turns out that if the observed gas motions are isotropic, the kinetic pressure support is less than 10\% of the thermal pressure support in the cluster core. The assumption of hydrostatic equilibrium in cluster mass measurements is thus likely not violated even in the dynamically active AGN hosting cluster cores. 

Based on the analysis of surface brightness fluctuations measured with {\it Chandra}, \citet{zhuravleva2014} performed indirect estimates of gas motions in the Perseus cluster even before the {\it Hitomi} observations. These estimates turned out to be  broadly consistent with the SXS measurements.  \citet{zhuravleva2014} showed that the heating rate from the dissipation of gas motions might actually be capable of balancing the radiative cooling at each radius in the Perseus cluster. 

Prior to the {\it Hitomi} observation of the Perseus cluster, the velocity broadening of emission lines has only been observed in giant ellipticals, groups, and clusters with strongly centrally peaked surface brightness distributions, using the reflection grating spectrometers (RGS) on board {\it XMM-Newton} \citep{sanders2011,sanders2013,pinto2015}. For more than half of the systems, the measurements of line widths provide a 68\% upper limit of 200 km s$^{-1}$. Because RGS is a slit-less dispersive spectrometer, the spectral lines observed by this instrument are also broadened by the spatial extent of the source, significantly limiting the velocity broadening measurements, plaguing them by a large systematic uncertainty. 

However, for the low temperature X-ray atmospheres of groups and giant ellipticals with $kT\lesssim0.9$~keV, the grating spectrometers on {\it XMM-Newton} enable indirect velocity measurements using the effect of resonance scattering \citep{gilfanov1987,churazov2010b}. While the hot atmospheres are mostly optically thin, at the energies of certain transitions resonance scattering may make the gas optically thick. The most affected by resonance scattering is the \ion{Fe}{xvii} line at 15.01~\AA. The optical depth (the expected suppression of the line) {\it decreases} with the increasing small scale turbulence. Conveniently, the neighbouring unresolved blend of the same ion at 17.05 and 17.10~\AA\ is optically thin and virtually unaffected by resonance scattering. Therefore, the comparison of their intensities allows us to measure the magnitude of resonant scattering and estimate the characteristic velocities of small scale turbulence. \citet{werner2009} and \citet{deplaa2012} measured the suppression of the 15.01 \AA\ \ion{Fe}{xvii} line in a sample of X-ray bright nearby galaxies and found that the turbulent pressure support could span from less than $\sim5$\% to over 40\% of the thermal pressure in these systems. \citet{ogorzalek2017} combined the  {\it XMM-Newton} RGS measurements of both line broadening \citep[based on][]{pinto2015} and resonant scattering for 13 galaxies and found a mean non-thermal pressure support of $\sim 6$\%. This non-thermal pressure support is consistent with the {\it Hitomi} SXS measurements of the Perseus cluster and supports the picture of a quasi-continuous gentle AGN feedback.  

The velocities of the cold and warm gas phases within the hot galactic, group and cluster atmospheres might also reflect the velocities in the hot ICM \citep[e.g.][]{gaspari2018}. Sub-mm, infrared, and optical spectra of the line emitting nebulae might thus also provide indirect estimates on the velocities in hot galactic atmospheres. 

Most of these velocity measurements were performed in relaxed systems, which are currently not undergoing major mergers. Mergers produce large bulk motions as well as in strong turbulence. They also result in weak shocks and cold fronts which are discussed in detail in \citet{markevitch2007}.

\section{The chemical composition of hot atmospheres}
\label{sec:5}

The first X-ray spectra of clusters of galaxies, obtained in the late 1970s \citep{mitchell1976,mushotzky1978}, revealed the presence of K-shell Fe line emission. Few years later, emission lines of other metals were also detected by the {\it Einstein} satellite \citep{lea1982}. These observations provided the very first evidence that the ICM has been significantly enriched with metals\footnote{All elements heavier than H and He are usually called ``metals'' in astrophysics.}. X-ray spectroscopy is a particularly powerful tool for studying the chemical enrichment, because all important products of AGB stars (C and N) and supernovae (O, Ne, Mg, Si, S, Ar, Ca, Cr, Mn, Fe, and Ni) have their K-shell lines in the X-ray band (as well as the L-shell lines of Si, S, Ar, Ca, Fe, and Ni, with the Fe-L lines being particularly prominent). Compared to the abundances measured in other objects at longer wavelengths (e.g. stellar abundances or the chemical composition of optical emission line nebulae), the X-ray measurements of the abundances in the hot ICM are remarkably accurate. Unlike the convention in the field, the abundances measured via X-ray spectroscopy can easily be published on a linear scale and the accuracy is often mainly limited by our knowledge of the atomic data \citep[e.g. see][]{hitomi2018_atomic}.

The most important questions addressed by the studies of the chemical composition of atmospheric gas include:
\begin{enumerate}
 \item How does the chemical composition of the atmospheric gas compare with the chemical composition of stars?
 \item What are the astrophysical sources of the enrichment? What is the fraction of Type Ia (SNIa; i.e. a thermonuclear explosion of a white dwarf when it reached the Chandrasekhar limit of 1.4~$M_{\odot}$ following a mass transfer from a stellar companion or a collision with another white dwarf) to core-collapse supernovae (SNcc; i.e. a gravitational collapse of a massive star at the end of its life leading to a supernova explosion) contributing to the enrichment of the intergalactic medium? 
 \item What can we learn about supernova nucleosynthesis from the chemical abundance studies of hot atmospheres? 
 \item At which cosmic epoch were the metals ejected from the stars and galaxies? Was it before or after the epoch of cluster formation?
 \item What are the main transport mechanisms that drive the metals out of galaxies and spread them across the intergalactic medium? 
\end{enumerate}{}

\begin{figure}
\includegraphics[scale=.33]{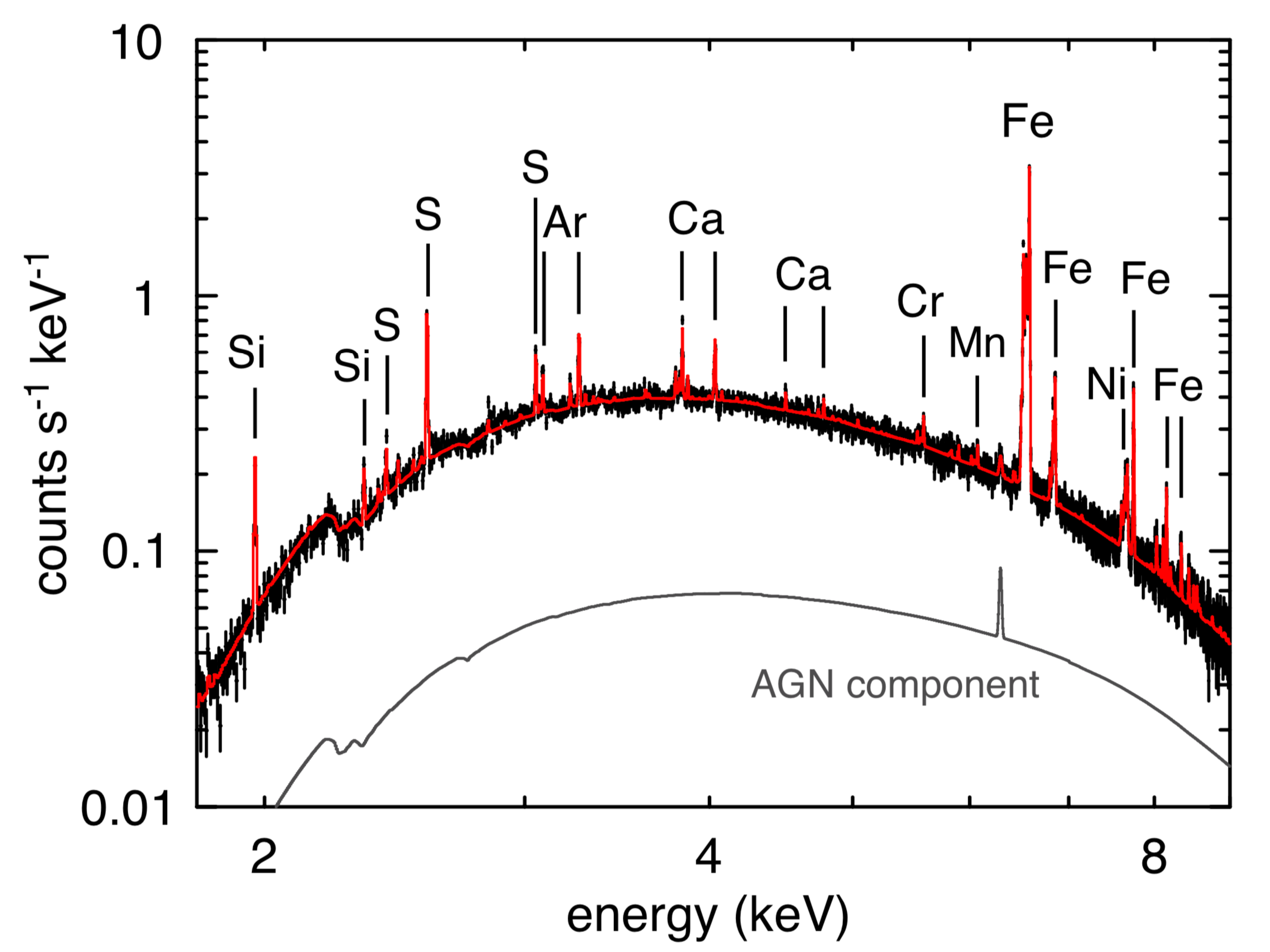}
\caption{The {\it Hitomi} SXS spectrum of the Perseus cluster resolves the emission lines of eight different elements produced by Type Ia and core-collapse supernovae. Its analysis revealed that the relative abundances of elements produced by supernovae are very similar to the relative abundances of the same metals in the Sun \citep{hitomi2017}.}
\label{hitomispec}    
\end{figure}

Addressing the first three questions requires the knowledge of the detailed chemical composition of the atmospheric gas, which means measurements of the abundances of as many supernova products as possible. The unique capabilities of \textit{XMM-Newton} and, more recently \textit{Hitomi} (see Fig.~\ref{hitomispec}), revealed that the relative abundances of the observed metals are consistent with those in the Sun \citep{deplaa2007,mernier2016a,hitomi2017,mernier2018b,simionescu2019}. This means that, similarly to the chemical enrichment models of our own Galaxy \citep[for a review, see][]{nomoto2013}, the enrichment of hot atmospheres has been dominated by only two types of supernovae: SNIa and SNcc. It seems that the contribution of population III stars (i.e. the very first stars in the Universe, made entirely of primordial hydrogen and helium) was negligible \citep[e.g.][]{werner2006}. 

Finding similar abundance ratios in our Solar neighbourhood and in the hot atmospheres of the largest objects in the Universe is surprising. The chemical composition of cluster atmospheres, which have been enriched by billions of supernovae exploding in thousands of galaxies should reflect the average chemical composition of the Universe. The result that the relative metal abundances of our Sun are so similar to those in clusters could be due to the fact that our Sun is an average star in a kind of galaxy, where most of the stars and metals in the Universe were formed. This result also indicates that the material from which our Sun formed has been a well mixed product of many supernovae.

The fourth question, about the epoch of the metal ejection from galaxies into the intergalactic gas, can be addressed in two different ways. The most direct approach is X-ray spectroscopy of galaxy clusters at various redshifts in order to look for an evolution of their chemical abundances with cosmic time. Although many studies have attempted to use this approach \citep[e.g.][]{balestra2007,anderson2009,baldi2012,ettori2015,mcdonald2016,mantz2017}, the very large uncertainties due to the low signal-to-noise at high redshifts makes it difficult to draw firm conclusions. Moreover, for the same reason this exercise is restricted to the brightest (i.e. most massive) clusters. Despite these limitations, most studies report essentially little or no cosmic evolution in the metallicity of hot atmospheres out to $z \simeq 1-1.5$, when the Universe was about half of its current age. 

\begin{figure}
\begin{center}
\includegraphics[scale=0.46]{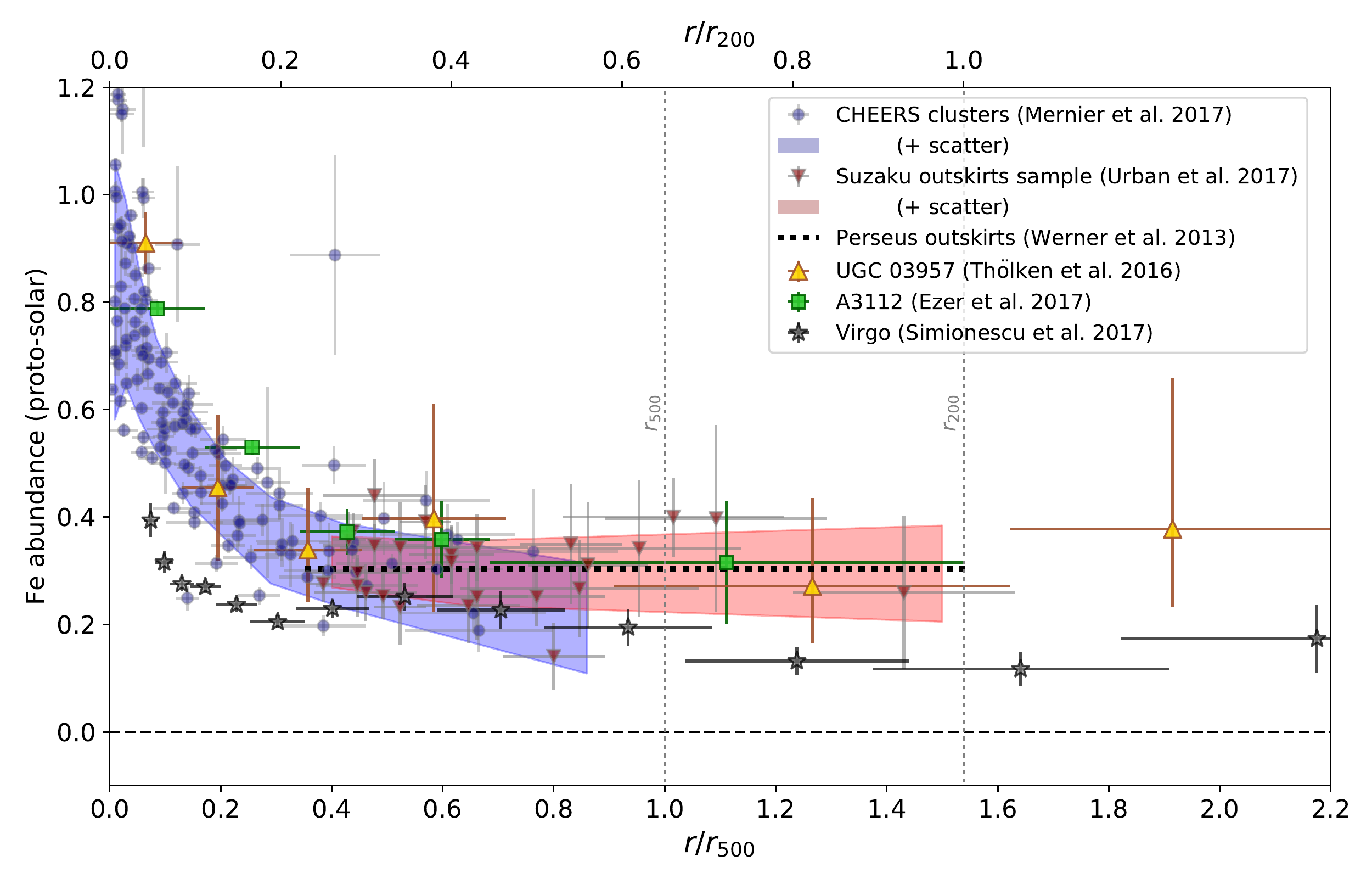}
\caption{Average radial Fe abundance profiles for cool-core clusters from various measurements in the literature \citep{mernier2018c}.}
\label{fig:Feprofile}
\end{center}
\end{figure}

The spatial distribution of metals in nearby systems also provides valuable information about the period when the bulk of the ICM enrichment took place (Fig. \ref{fig:Feprofile}). In particular, the \textit{Suzaku} X-ray satellite revealed that, instead of decreasing monotonically with radius, the Fe abundance in clusters outskirts remains remarkably uniform at $Z_{\rm Fe} \sim0.3$ Solar out to at least $r_{200}$ \citep{fujita2008,werner2013,urban2017}. According to simulations \citep[e.g.][]{biffi2018a,biffi2018c}, this flat distribution is the key signature of metals having escaped their galaxy hosts at $z \simeq 2-3$ (more than 10 billion years ago), i.e. \textit{before} these galaxies were incorporated into clusters. These results led to the so-called ``early enrichment scenario'', in which the bulk of metals has been ejected into the IGM before clusters started to form and grow. 

As shown in Fig.~\ref{fig:Feprofile}, relaxed cool-core clusters exhibit a central increase of Fe. Because these clusters host BCGs in their centres, this Fe peak was initially attributed to the stellar populations of these galaxies \citep{bohringer2004,degrandi2004}. While BCGs lack significant star formation and SNcc explosions, SNIa, due to their long delay times, may still keep enriching the BCGs with heavy metals, such as Fe and Ni. Lighter elements like O, Mg, and Si are mainly produced by SNcc and were thus not expected to be seen in large amounts within the central region. It was later found, however, that all elements, including O, Mg, and Si, exhibit a central abundance peak that follows that of Fe \citep{deplaa2006,simionescu2009b,million2011,mernier2017}. This surprising discovery indicates that stars that are seen today in BCGs have little to do with the enrichment of the hot gas pervading them. In other words, the bulk of the gas enrichment must have occurred early on, \textit{before the hot medium ended up in the BCG}, during or perhaps even before the formation of the cluster. This conclusion is also supported by the observation that while the stellar populations of BCGs have super-Solar $\alpha$/Fe abundance ratios \citep[ratios of SNcc products over iron, which is produced mainly by SNIa;][]{conroy2014}, their hot atmospheres have Solar $\alpha$/Fe ratios. The different composition points to a different origin of metals. 

\begin{figure}
\begin{center}
\includegraphics[scale=0.35]{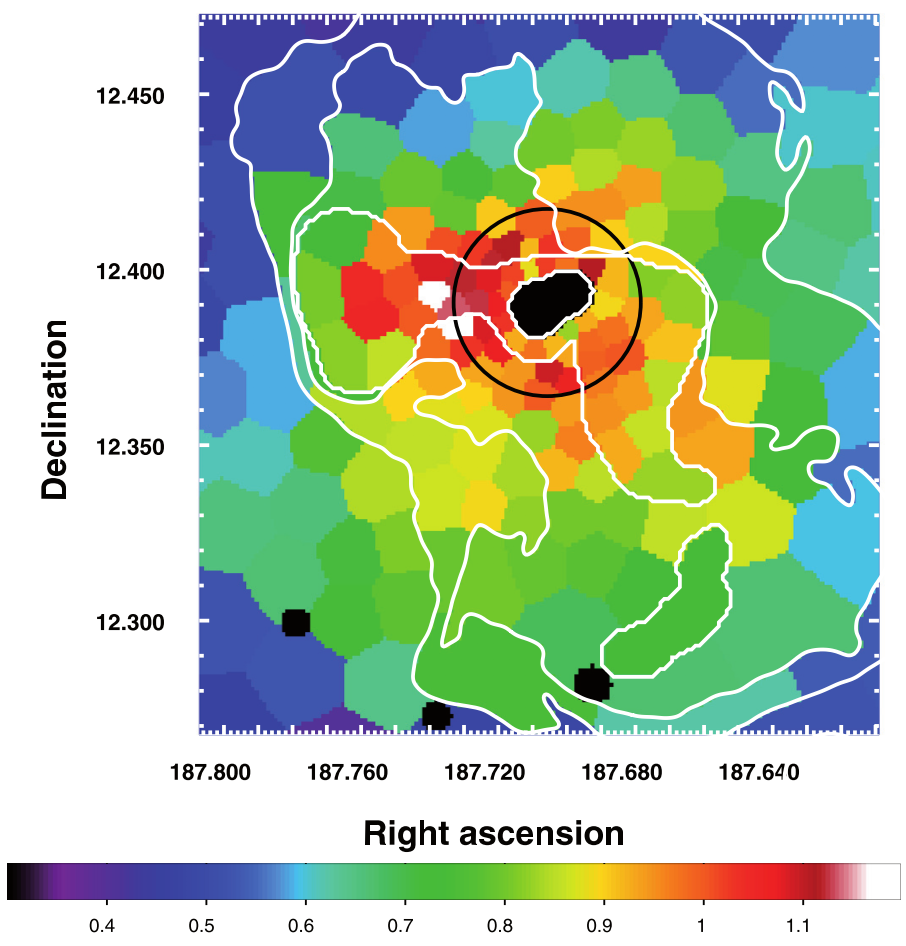} 
\includegraphics[scale=0.27]{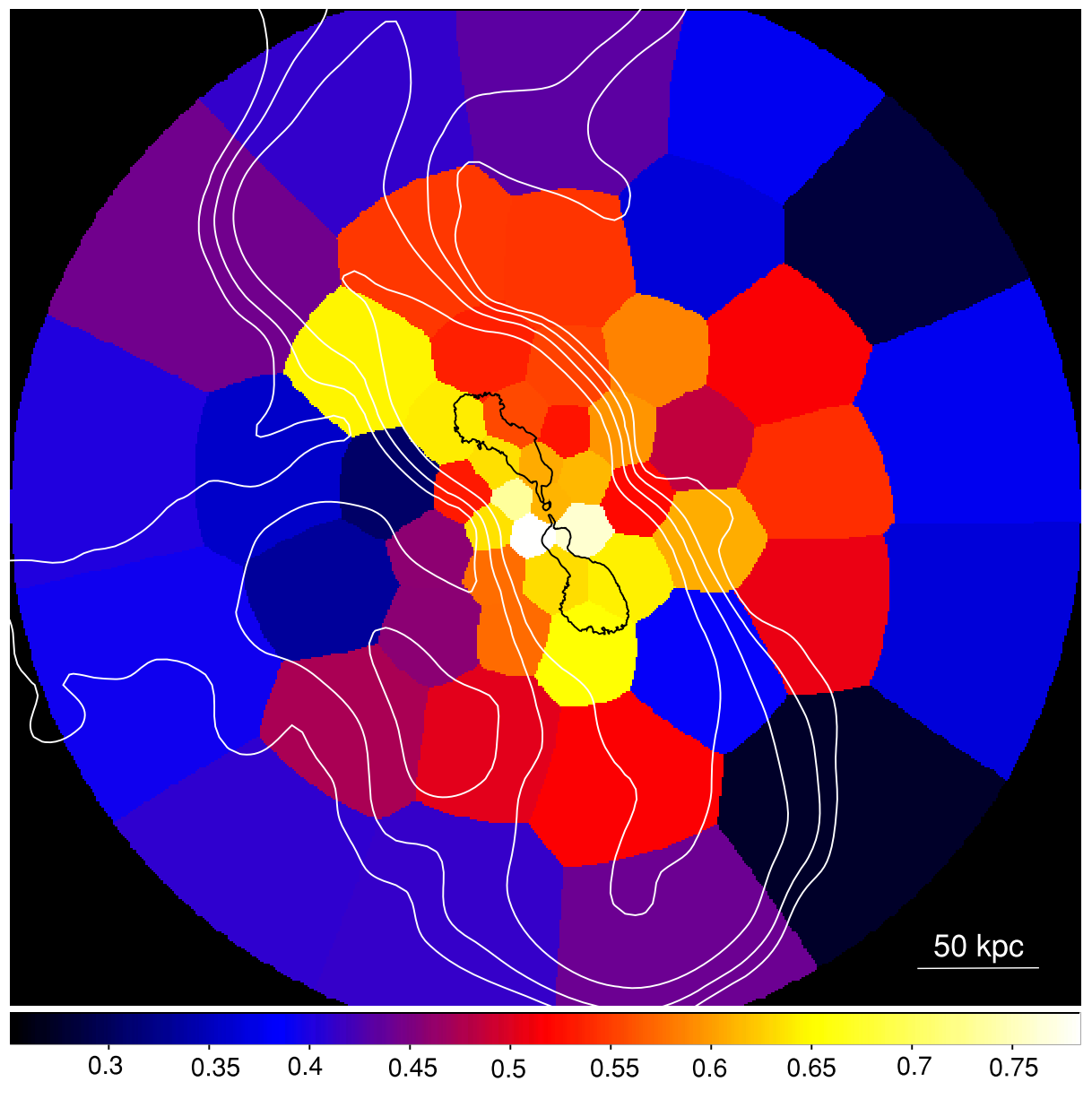} 
\caption{Projected Fe abundance maps for the hot atmospheres of M~87  \citep[\textit{left panel;}][]{simionescu2008} and the Hydra A cluster  \citep[\textit{right panel;}][]{kirkpatrick2009}. The 330 MHz and 1400 MHz radio contours, over-plotted in white and black, respectively, trace the jets from the central supermassive black hole.}
\label{fig:metalmaps}
\end{center}
\end{figure}

Interestingly, the spatial distribution of metals in hot atmospheres also provides crucial clues to answer our last question. Simulations which reproduce the observed flat metal distribution in clusters outskirts \citep[e.g.][]{biffi2018a,biffi2018c} indicate that the activity of supermassive black holes was necessary to eject the metals from their galactic potential before the formation of clusters. In some cases, a gentle uplift of metals by AGN feedback can be seen directly around the central dominant galaxies of nearby clusters, where metal-rich gas is observed along the radio jets of the central AGN and around their cavities \citep{kirkpatrick2011}. The clearest observational evidence of uplift has been obtained for M\,87 \citep[][Fig. \ref{fig:metalmaps} left]{simionescu2008} and for the Hydra A \citep[][Fig. \ref{fig:metalmaps} right]{simionescu2009b,kirkpatrick2009}. In these cases, however, the deep potential well of such mature systems prevents metals to effectively diffuse out and to enrich their surroundings at very large scales. Ejection of metals by galaxies in proto-cluster environments (i.e. at $z \gtrsim$ 2) via their central supermassive black holes is yet to be observed directly. 

In addition to AGN feedback, other processes may play a significant role in redistributing metals within hot atmospheres \citep[for a review, see][]{schindler2008}. These are: (i) ram-pressure stripping (i.e. when a metal-rich galaxy falls into a galaxy cluster, the pressure of the pervading ICM strips its local interstellar medium and a tail of metal-rich gas forms behind the galaxy), (ii) sloshing (i.e. when the ICM is slightly displaced from its potential well, circular motions start to take place and may gently transport metals on large scales), and (iii) galaxy-galaxy interactions. In addition, clusters of galaxies contain a significant fraction of intracluster stars, which do not belong to any galaxy. For this reason, they can, in principle, release metals into the hot atmosphere more efficiently than stars that reside in a strong galactic potential. 
Currently, the exact contribution of all these mechanisms is not well known, and future dedicated observations and simulations will be needed to clarify the relative contribution of these processes.

\section{Future studies of hot atmospheres}
\label{sec:7}

The current generation of X-ray satellites (\textit{Chandra}, \textit{XMM-Newton}, and \textit{Suzaku}) revolutionized our understanding of hot atmospheres pervading massive galaxies, groups and clusters. These missions greatly benefited from the all sky survey performed by the {\it ROSAT} satellite launched in 1990. The {\it ROSAT} survey was crucial for the mapping of the X-ray sky and discovering new systems, yet it still remains unsurpassed.

Before the launch of the major next generation X-ray observatories it will be critically important to perform a new, more sensitive sky survey. That is the objective of the eROSITA instrument onboard the recently launched Russian-German \textit{Spectrum-Roentgen-Gamma} (\textit{SRG}) mission, which is currently mapping the X-ray sky with a much improved spatial resolution and sensitivity \citep{merloni2012}. This will allow the discovery of approximately 50 000-100 000 new galaxy clusters and about 3 millions new AGN. 

Perhaps the most significant improvements in the capabilities of the future major X-ray observatories are expected in spectral resolution. They will allow us to probe directly small-scale turbulence together with large-scale gas motions. The high spectral resolution also allows us to access more metal lines, providing increasingly accurate elemental abundance measurements. Micro-calorimeters, which convert a tiny change of temperature due to the incoming X-ray photon into photon energy, have demonstrated their potential by the spectacular results from the \textit{Hitomi} SXS observation of the Perseus cluster. A re-flight of this Japanese mission called \textit{XRISM} \citep[X-Ray Imaging and Spectroscopy Mission;][]{tashiro2018}, is planned for the year 2022 and its key micro-calorimeter instrument Resolve will routinely provide X-ray spectra with an energy resolution of 5~eV. 

\begin{figure}
\includegraphics[scale=.75]{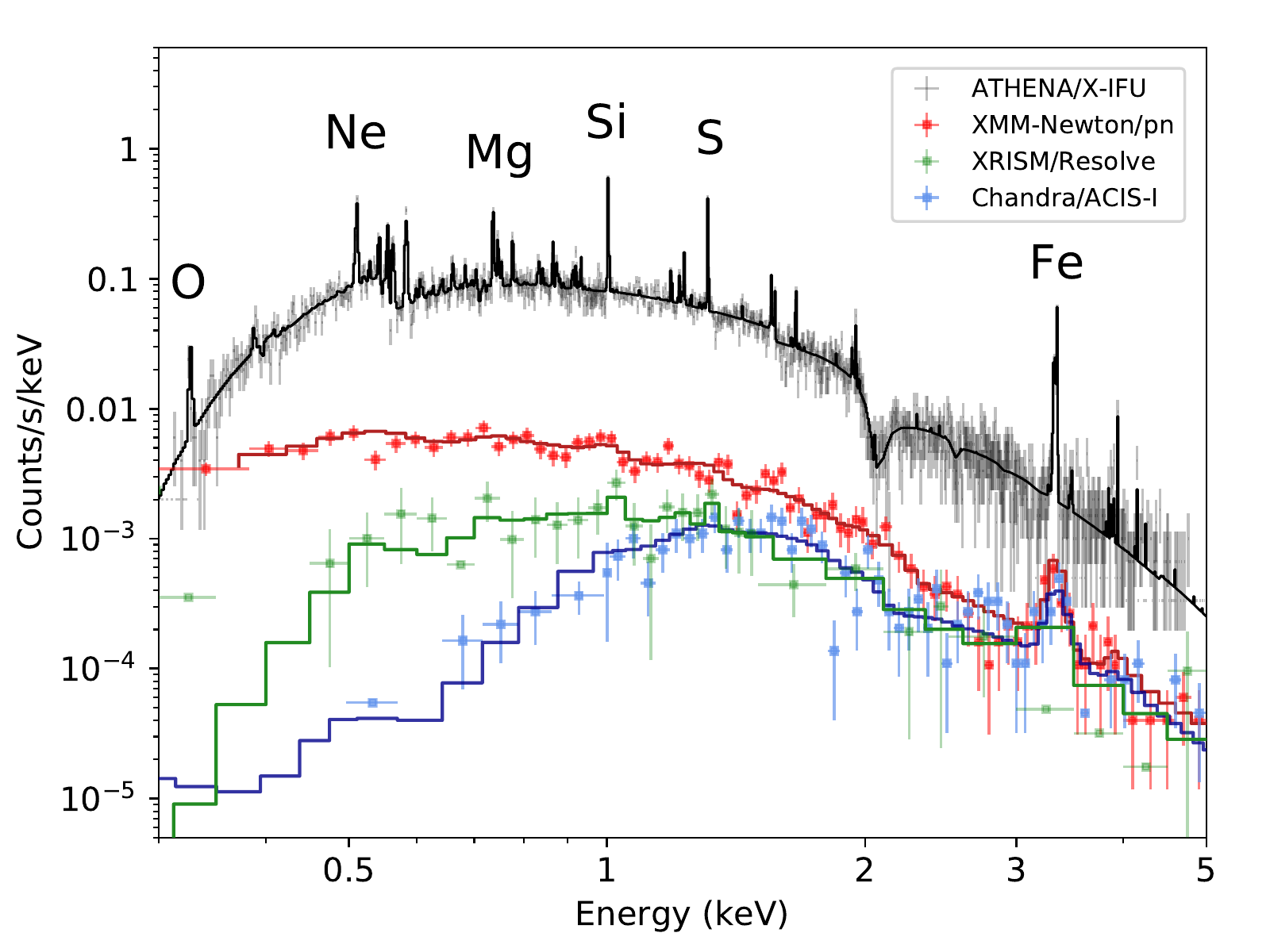}
\caption{A simulated {\it Athena} X-IFU spectrum of a $kT=3$~keV galaxy cluster at a redshift of $z=1$ obtained assuming a deep 250~thousand second (about three days) long observation. For comparison, simulated observations, assuming equally long exposure times, are shown for {\it XMM-Newton} EPIC/pn, {\it Chandra} ACIS-I, and {\it XRISM} Resolve.}
\label{fig:athena}   
\end{figure}

Later on around 2031, ESA's next generation X-ray observatory \textit{Athena} (\textit{Advanced Telescope for High-ENergy Astrophysics}) is expected to push our knowledge of hot atmospheres to yet another level. \textit{Athena} will be equipped with a cryogenic X-ray spectrometer, based on a large array of Transition Edge Sensors (TES), offering 2.5 eV spectral resolution, with 5 arcsec pixels, over a field of view of 5 arcminutes in diameter \citep[the X-IFU instrument;][]{barret2018}. Moreover, its effective area is expected to be an order of magnitude larger than that of \textit{XRISM} (see Fig.~\ref{fig:athena}). This unprecedented sensitivity will be of vital importance as it will allow to probe high-redshift systems and provide measurements to constrain the evolution of the thermodynamic properties and metallicities of hot atmospheres. \textit{Athena} will be able to detect groups at early epochs (out to look-back times of $\sim$10 Gyr), which will be essential for understanding the formation of large scale structures.

The high spectral resolution is also necessary to survey the distribution and properties of the faint circumgalactic and intergalactic medium permeating the filaments of the cosmic web. This will be the primary aim of the future Chinese \textit{HUBS} (\textit{Hot Universe Baryon Surveyor}) mission\footnote{http://hubs.phys.tsinghua.edu.cn/en/index.html}. \textit{HUBS} will combine a large field-of-view with an excellent spectral resolution in order to map the emission and absorption in the \ion{O}{vii} and \ion{O}{viii} lines. These measurements will be crucial to detect and ``map'' the bulk of the ordinary matter in our nearby Universe. The yet to be approved Japanese \textit{SuperDIOS} mission \citep{yamada2018} will extend the capabilities of {\it HUBS}. Its 10 arcsec spatial resolution will allow a robust filtering of contaminating X-ray point sources and thus a more reliable census of diffuse warm-hot baryons.

The \textit{Athena} X-ray observatory is expected to achieve a 5 arcsec spatial resolution, which is significantly better than that of \textit{XMM-Newton}, but still worse than that of \textit{Chandra}. In particular the studies of galaxy scale atmospheres would strongly benefit from a high spatial resolution, which would allow us to resolve out the bright point sources in these systems. But also the studies of subtle features in nearby galaxy clusters (e.g. shocks, cold fronts, ripples, etc.) or substructures in high-redshift systems (cavities, signatures of sloshing, merging sub-clusters, etc.) will benefit from combining the unprecedented spectral resolution, large effective area, and excellent \textit{spatial} resolution. This is the ambitious goal of the proposed NASA mission \textit{Lynx}, which aims to combine the capabilities of \textit{Athena} with a \textit{Chandra}-like spatial resolution and add a high resolution grating spectrometer.

The post-\textit{Athena} (-\textit{Lynx}) future of X-ray observatories remains far and not clear. It depends strongly on the future science questions uncovered by the upcoming facilities and on future technology that is yet to be demonstrated. However, it is never too early to ``think further'' and propose a very long term view on astrophysical observations (e.g. see the Voyage 2050 call by ESA). Even though the currently planned missions are expected to reveal much more about the diffuse medium outside of the  gravitational potential wells of massive clusters, groups and galaxies, they will still only probe the tip of the iceberg. The physical properties of the majority of ordinary matter in the local Universe will remain unexplored even after {\it Athena, Lynx, HUBS}, and {\it Super-DIOS}. This task awaits for a dedicated mission with a large effective area, field-of-view, as well as high spatial and spectral resolution. Within about two decades, megapixel size TES detector arrays with sub-eV spectral resolution in the soft X-ray band are expected to be available. Our ambition and hope is, that such sizable high-resolution X-ray spectrometer arrays combined with large X-ray mirrors, that are currently under development, will finally enable us to build the \textit{Cosmic Web Explorer} mission \citep{simionescu2050}, which will survey the physical properties of most of the medium permeating the cosmic ocean.

\begin{acknowledgement}
This work was supported by the Lend\"{u}let LP2016-11 grant awarded by the Hungarian Academy of Sciences.\end{acknowledgement}

\bibliographystyle{aps-nameyear}
\bibliography{clusters.bib}

\end{document}